\documentclass[aps,prd,twocolumn,10pt,showkeys,superscriptaddress,nobibnotes,longbibliography]{revtex4-1}

\usepackage{mathrsfs,graphicx,rotating,amsmath,amsfonts,mathtools,booktabs,amssymb,wasysym}
\usepackage{hyperref}
\usepackage[table,xcdraw,dvipsnames]{xcolor}
\usepackage{subfig}
\usepackage{caption}
\hypersetup{
     colorlinks   = true,
     citecolor    = blue,
     urlcolor     = blue,
     linkcolor    = blue
}

\newcommand{\BR}{\hbox{BR}}

\newcommand{\med}[1]{\langle #1\rangle}

\newcommand{\MDM}{M_{\text{DM}}} 
\newcommand{\Q}{{\mathcal{Q}}}

\newcommand{\be}{\begin{equation}}
\newcommand{\ee}{\end{equation}}

\newcommand{\PRL}{Phys. Rev. Lett.}

\makeatletter

\def\hhref#1{\href{http://arxiv.org/abs/#1}{arXiv:#1}}
\usepackage{xstring} 
\newcommand{\hhrefq}[1]{\IfSubStr{#1}{:}{\href{http://inspirehep.net/search?ln=en&ln=en&p=#1&of=hb&action_search=Search&sf=&so=d&rm=&rg=25&sc=0}{InSpires:#1}}{\hhref{#1}}}

\def\art{\@ifnextchar[{\eart}{\oart}}
\def\eart[#1]#2#3#4#5#6{{\rm #2}, {\em #3 \bf #4} {\rm (#6) #5} ({\em #1})}
\def\article{\@ifnextchar[{\earticle}{\oarticle}}
\def\oarticle#1#2#3#4#5#6{{\rm #1}, {``#6''}, {\rm #2 #3 (#5) #4}}
\def\earticle[#1]#2#3#4#5#6#7{{\rm #2}, {``#7''}, {\rm #3 #4 (#6) #5}  [\hhrefq{#1}]}
\def\hepart[#1]#2{{\rm #2, \sl#1}}
\def\heparticle[#1]#2#3{#2, { ``#3''} [\hhrefq{#1}]}

\begin{document}
{\hfill CP3-Origins-2019-17 DNRF90\hfill}

\title{TeV-Scale Thermal WIMPs: Unitarity and its Consequences}

\author{Juri Smirnov}
\thanks{{\scriptsize Email}: \href{mailto:smirnov@cp3.sdu.dk}{smirnov@cp3.sdu.dk}; {\scriptsize ORCID}: \href{http://orcid.org/0000-0002-3082-0929}{ 0000-0002-3082-0929}}
\affiliation{$\text{CP}^3$-Origins, University of Southern Denmark}	

\author{John F. Beacom }
\thanks{{\scriptsize Email}: \href{mailto:beacom.7@osu.edu}{beacom.7@osu.edu}; {\scriptsize ORCID}: \href{http://orcid.org/0000-0002-0005-2631}{0000-0002-0005-2631}}
\affiliation{Center for Cosmology and AstroParticle Physics (CCAPP), The Ohio State University, Columbus, OH 43210, USA}
\affiliation{Department of Physics, The Ohio State University, Columbus, OH 43210, USA}
\affiliation{Department of Astronomy, The Ohio State University, Columbus, OH 43210, USA}

\date{\today}

\begin{abstract}
We re-examine unitarity bounds on the annihilation cross section of thermal-WIMP dark matter.  For high-mass pointlike dark matter, it is generic to form WIMP bound states, which, together with Sommerfeld enhancement, affects the relic abundance.  We show that these effects lower the unitarity bound from 139 TeV to below 100 TeV for non-self-conjugate dark matter and  from 195 TeV (the oft-quoted value of 340 TeV assumes $\Omega_{DM} h^2 = 1$)  to 140 TeV for the self-conjugate case.  For composite dark matter, for which the unitarity limit on the radius was thought to be mass-independent, we show that the largest allowed mass is 1 PeV.  In addition, we find important new effects for annihilation in the late universe.  For example, while the production of high-energy light fermions in WIMP annihilation is suppressed by helicity, we show that bound-state formation changes this.  Coupled with rapidly improving experimental sensitivity to TeV-range gamma rays, cosmic rays, and neutrinos, our results give new hope to attack the thermal-WIMP mass range from the high-mass end. 
\end{abstract}  

\maketitle

%%%%%%%%%%%%%%%%%%%%%%%%%%%%%%%%%%%%%%%%%%%%%%%%%%%%%%%
%%%%%%%%%%%%%%%%%%%%%%%%%%%%%%%%%%%%%%%%%%%%%%%%%%%%%%%

\section{Introduction}

The unknown particle nature of dark matter has inspired a plethora of imaginative models~\cite{Ge:2019voa, Berges:2019dgr, Blennow:2019fhy, Arcadi:2019lka,Cirelli:2018iax,1705.05567}.  Among them, one well-motivated model is unique in its simplicity and specificity, and that is a thermal-relic WIMP that annihilates to Standard-Model particles~\cite{0404175, Steigman:1984ac, Arcadi:2017kky,Roszkowski:2017nbc}.  While this may not be the correct description of nature, it is essential that this hypothesis be fully tested.  This is challenging but possible.

In this model, the early universe annihilation rate factor is determined from the dark matter relic abundance as $\langle \sigma v \rangle  = (2.2 \times 10^{-26} \, \text{cm}^3 \, \text{s}^{-1}) (0.12 / \Omega_{DM} h^2)$, where this is the total cross section to all final states~\cite{SteigmanDasguptaBeacom}.  (We quote the value at large dark-matter masses; at smaller masses, it is larger.)  If annihilation proceeds through $s$-wave scattering, as is well motivated, then the late-universe annihilation rate factor is the same.  Given the density distribution of dark matter, determined by gravitational probes~\cite{Aghanim:2018eyx}, upper limits on the fluxes of energetic particles (gamma rays, cosmic rays, and neutrinos) then determine upper limits on $\langle \sigma v \rangle_i$, where $i$ denotes particular final states.  Importantly, in a generic thermal-WIMP model, only the total annihilation rate factor is model-independent; those to particular final states --- as well as production rates at colliders~\cite{DM@LHC, Baltz:2006fm, Han:2018wus} and elastic-scattering rates in underground detectors~\cite{Goodman:1984dc, Cushman:2013zza, Schumann:2019eaa} --- are model-dependent.

Fully testing the thermal-WIMP hypothesis requires reaching a cross section sensitivity below this prediction, and doing so for all WIMP masses.   Moreover, though searches for annihilation products test the partial cross sections to particular final states, one must combine those to determine the upper limit on the cross section to all Standard-Model final states~\cite{UpperBoundCross, LeaneSlatyerBeacom}.  If one excludes neutrinos from being dominant final states, then the lower limit is $\simeq 20$ GeV~\cite{LeaneSlatyerBeacom}, to be contrasted with the limits claimed assuming favorable final states, which approach $\simeq 100$ GeV~\cite{FermiDwarf}.  (WIMPs that annihilate {\it only} to neutrinos are excluded below $\sim 10$ MeV by big bang nucleosynthesis~\cite{NollettSteigman1, NollettSteigman2}; if there is {\it any} annihilation to other channels, the mass limit from the cosmic microwave background is much larger~\cite{LeaneSlatyerBeacom, Galli:2013dna, Padmanabhan:2005es}.)  It is well known that improvements in the sensitivity of existing searches are possible, and that, if no signals are found, the thermal-relic mass range will be progressively attacked from the low-mass end~\cite{1408.4131}.

But what about the high-mass end?  The largest allowed mass of a thermal WIMP is then determined only by the theoretical bound from $s$-wave unitarity, and is $\simeq 340$ TeV  (for $\Omega_{DM} h^2= 1$; see below for $\Omega_{DM} h^2 < 1$)  for pointlike dark matter~\cite{GriestKamionkowski}.  For composite dark matter, the lower limit on the radius is $7.5 \times 10^{-7} \, \rm fm$, independent of mass~\cite{GriestKamionkowski}.  In 1990, when these limits were set, the experimental sensitivity to high-mass dark matter annihilation was vastly inadequate.  Today, it is much better but still inadequate, though that will change due to new generations of experiments and better understandings of astrophysical backgrounds.  This prompts a re-examination of thermal WIMPs at the largest masses.  New theoretical developments~\cite{Nima, Cassel, Petraki, Wise, CosmoBS} make this especially relevant.

Saturating unitarity requires a large cross section.  In the minimal scenario, we consider, where the only new particle is the WIMP, the options to achieve this are limited.  (For example, it is not possible to invoke resonances~\cite{Pavel, DiracDM}, as these require additional particles to be the mediators.)  A key insight is that when the WIMP mass is very large compared to all Standard-Model particles, there are only light mediators, which induce long-range potentials and enhance cross sections through the Sommerfeld effect~\cite{Nima, Cassel, Hisano:2003ec, Hisano:2004ds, Hisano:2006nn, 0812.0559}. Including near threshold effects the Sommerfeld effect was further investigated in~\cite{1609.00474, 1706.01894}. (The dark matter itself can not serve as a mediator, as this topology would unavoidably open a decay channel to Standard Model particles.)  Further, based on more recent work, there are also generically bound-state effects~\cite{Petraki, CosmoBS}.  Only by taking all of these effects into account, as we do in this paper, can accurate results be obtained.  Further, these effects change the prospects for the detection of late-universe annihilation products.

This paper is structured as follows. In Sec.~\ref{sec:perturbative}, we review bound-state effects on WIMP freeze-out and how these affect the unitarity bound.  In Sec.~\ref{sec:model}, we calculate the bound-state effects on the unitarity bound for pointlike dark matter and calculate the largest allowed WIMP masses.  In Sec.~\ref{sec:extended}, we extend our calculations to composite dark-matter states of finite size.  In Sec.~\ref{sec:conclusions}, we summarize our results, emphasizing the path towards experimental sensitivity at high masses overtaking the unitarity bounds for the first time.  The ultimate goal is to test thermal WIMPs over the full mass range by attacking from both the low-mass and high-mass ends.

%%%%%%%%%%%%%%%%%%%%%%%%%%%%%%%%%%%%%%%%%%%%%%%%%%%%%%%%
%%%%%%%%%%%%%%%%%%%%%%%%%%%%%%%%%%%%%%%%%%%%%%%%%%%%%%%%

\section{Overview of Bound-State Effects for Heavy Dark Matter}
\label{sec:perturbative}

In this section, we briefly review previous studies of bound-state effects on annihilation, define in what way the present framework is different, and note the implications for freeze-out.

%%%%%%%%%%%%%%%%%%%%%%%%%%%%%%%%%%%%%%%%%%%%%%%%%%%%%%%%

\subsection{Bound-State Effects on Freezeout}

Bound-state effects on dark matter (DM) annihilation have been discussed only recently, and only in certain cases.  In the context of a $U(1)$ model, they were found in Ref.~\cite{Petraki} to be important, though Ref.~\cite{Wise} argued the opposite.   A follow-up study \cite{PetrakiNew} discusses bound-state formation as a possibility to reach the unitarity bound in a perturbative abelian model, with a new hypothetical force carrier and a coupling strength close to its non-perturbative value.  Bound-state effects in a weakly interacting DM system were studied in Refs.~\cite{Slatyer, 0901.2125, 0812.0559}. However, only late time annihilation was considered and not the freeze-out process.  A consistent framework for an effective theory approach to the weak triplet (Wino) scenario was developed in Refs.~\cite{Braaten1, Braaten2, Braaten3}.  While models like these, or those with a new force carrier, such as a new neutral gauge boson suggested in Ref.~\cite{PranNath}, are certainly appealing, there are ways to introduce a more minimal dark sector with only one additional particle. 

In this work, we consider the simplest set-up for a dark sector, with one new particle charged under the weak force ($SU(2)_L$). Despite the existence of multiple components of different charge in the multiplet, there are only two free parameters, the particle mass and its representation under $SU(2)_L$.  As will become evident, the representation of the particle is its effective charge and determines its coupling strength to the weak gauge bosons. We show below that larger representations are extremely relevant for DM systems close to the unitarity bound. An important aspect arises from the fact that the weak force is non-abelian and thus, given two particles in a representation $R$  and $\bar{R}$, there is always an attractive (singlet) channel that makes Sommerfeld enhancement and bound-state formation absolutely generic phenomena.

The claim of Ref.~\cite{Wise} (see their Sec.~4), that bound state dissociation makes the net effect on the freeze-out irrelevant, was refuted in Refs.~\cite{CosmoBS, PetrakiNew}. This was done through developing an analytic asymptotic solution to the system of Boltzmann equations arising from all allowed bound states~\cite{Ellis:2015vaa, CosmoBS}. We use that mathematical framework to show the unavoidable effects of bound-state formation on the unitarity bound.  

Freeze-out in the presence of bound states is described by a set of coupled Boltzmann equations, one for the dark-matter number density and one for each allowed bound state $I$ with the formation cross section $\med{\sigma_I v_{\rm rel}}$ and annihilation rate $\Gamma_I$. The analysis relies on the fact that the DM bound state is not stable, as for example positronium, and can annihilate to SM particles or dissociate into WIMP constituents.  The crucial simplification follows from the fact that for the WIMP bound states we study, their lifetimes are much less than the Hubble time.  Then all the Boltzmann equations for bound states can be treated algebraically, neglecting the time derivatives, and inserted in the Boltzmann equation for the DM abundance \cite{CosmoBS}. This leads to an effective cross section for DM, where the bound states lead to new channels with temperature-dependent branching ratios into SM particles. Intuitively, this effective cross section takes into account the fact that a bound state can be broken by the plasma before it annihilates, and thus not affect the total DM number density. The breaking rate is related to the formation rate in equilibrium by the Milne relation $\Gamma_{\rm break} \, n_I  = \med{\sigma_I v_{\rm rel}} n_{\rm DM}^2$. Once the temperature drops far below the bound-state binding energy, the breaking is strongly suppressed. 

Quantitatively, this treatment leads to a single Boltzmann equation for $Y=Y_{\rm DM} = n_{\rm DM}/s$:
\begin{align}
\frac{dY}{dz} = - \frac{\lambda \, \med{ \sigma_{\rm eff}  v_{\rm rel} } (z)} {z^2} (Y^2 - Y_{\rm eq}^2),
\end{align}
where $\lambda =  \sqrt{\frac{g_{\rm SM}\pi}{45}} M_{\rm Pl} \MDM$, $z = M_{\rm DM}/T$   and 
\begin{align}
\med{\sigma_{\rm eff}  v}  =  \med{\sigma_{\rm ann}  v_{\rm rel}} +  \sum_{I=1}^\infty \med{\sigma_{\rm I}  v_{\rm rel}}  \BR(\rm B_I \rightarrow \rm SM), 
\end{align}
\begin{align}
\label{eqn:Braching}
& \text{with   } \quad  \BR(\rm B_i \rightarrow \rm SM) =    \frac{\Gamma_{\rm ann}}{\Gamma_{\rm ann} + \Gamma_{\rm break}}   \nonumber \\
& = \left[1 + \frac{\langle \sigma_{\rm I} v_{\rm rel} \rangle g_{\chi}^2 \,\MDM^3 \,  e^{- z\, E_{B_I}/\MDM}}{2 g_I\, (4 \pi z)^{3/2} \,  \Gamma_{\rm ann}}  \right]^{-1}\,,
\end{align}
where the rate for breaking of bound states follows from the Milne relation. 

This equation can be easily integrated numerically, but has an analytic asymptotic solution, which agrees very well with the numerical treatment,
\begin{align}
Y_{\rm DM}({\infty}) =  \frac{1}{ \lambda } \left(  \int_{z_f}^\infty \frac{ \med{\sigma_{\rm eff}  v_{\rm rel}} (z)}{z^2} dz   +   \frac{   \med{\sigma_{\rm eff}  v_{\rm rel}}  (z_f)}{z_f^2} \right)^{-1} \,,
\end{align}
with the inverse temperature at freeze-out $ \MDM/T_f =z_f$ given by the transcendental equation
\begin{align}
z_f&  = \ln{\left( \frac{2  g_\chi  \med{\sigma_{\rm eff}  v_{\rm rel}} (z_f) \lambda}{  (2  \pi z_f)^{3/2}} \right)}\,.
\end{align}
For multi-TeV DM, $z_f \approx 25$ is typical.

\begin{figure}[t]
\includegraphics[width=0.45\textwidth]{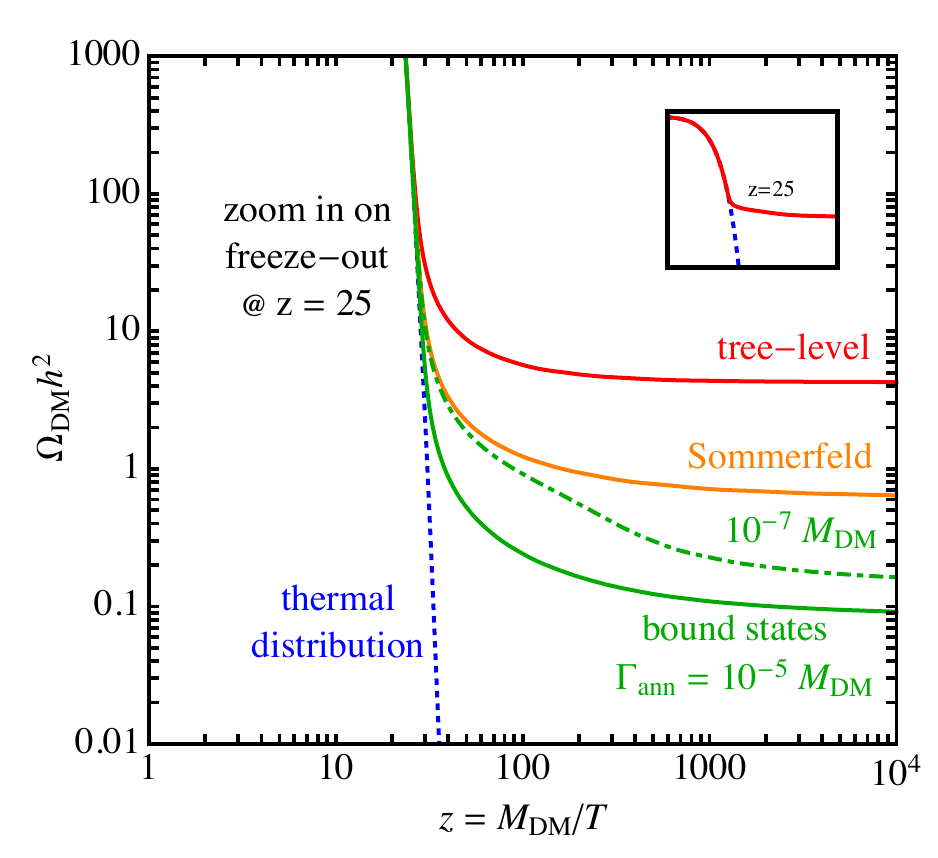}
\caption{\label{fig:FreezeOutBS} Effects on freeze-out due to the Sommerfeld effect alone and the additional effects of bound-state formation.  The inset shows the qualitative behavior at the time of deviation from the thermal DM abundance. Note in particular, that the DM depletion due to bound-state formation (green lines) sets in at later times than the Sommerfeld enhanced freeze-out. In particular in the case indicated by the dot-dashed green line, where the smaller bound-state annihilation rate of $\Gamma_{\rm ann} \approx 10^{-7} \MDM$ leads to a belated annihilation. This is a direct consequence of the branching ratio introduced in Eq.~(\ref{eqn:Braching}).}
\end{figure}

In Fig.~\ref{fig:FreezeOutBS}, we demonstrate the effects of the Sommerfeld enhancement and bound-state formation on freeze-out. Including the Sommerfeld effect leads to additional attraction among WIMPs and enhances the annihilation rate, which in turn reduces the relic abundance by $\mathcal{O}(10)$. The consideration of bound states is an additional effective annihilation channel and leads to a further $\mathcal{O}(10)$ reduction. This is not surprising since it is known that in the SM non-relativistic $e^+$ $e^-$ annihilation is dominated by positronium formation and its successive annihilation.  Additionally, the importance of the decay width of the considered bound state is highlighted. The typical annihilation width scales as $\alpha^5 M_{\rm DM}$, where $\alpha$ is the coupling strength of the interaction considered, and thus a typical width in a perturbative model would be of the order $10^{-5} M_{\rm DM} $ or smaller.  The observation we want to stress is, that while a bound state can be a reaction product of dark-matter interactions, its effect on the relic density strongly depends on its binding energy and decay width to SM particles. 

%%%%%%%%%%%%%%%%%%%%%%%%%%%%%%%%%%%%%%%%%%%%%%%%%%%%%%%%

\subsection{Effects on the Unitarity Bound}
\label{sec:unitarity}
As discussed in the classic paper of Griest and Kamionkowski~\cite{GriestKamionkowski}, conservation of probability limits the reaction cross section of DM annihilating to any final state for each partial wave by 
\begin{align}
\label{eq:unibound}
(\sigma v_{\rm rel})_{\rm total}^J < (\sigma v)_{\rm max} ^J= \frac{4\pi (2\,J+1) }{\MDM^2 v_{\rm rel}} \,.
\end{align}
Note the scaling of the bound with $v_{\rm rel}^{-1}$, which is not expected from contact type interactions, but is generic in the presence of long range forces. To understand the physical implications of the above inequality, we first discuss the cross sections that are relevant for the physical system.  In the following, $( \sigma v_{\rm rel})$ denotes non-averaged cross sections and $\med{\sigma v}$ denotes thermally averaged cross sections. The total (inelastic) reaction cross section is $(\sigma v_{\rm rel})_{\rm total}  = \sum_J (\sigma v_{\rm rel})_{\rm total}^J $. The total reaction cross section is composed of an annihilation part and the bound-state formation cross section $(\sigma v_{\rm rel})_{\rm total} = (\sigma v_{\rm rel})_{\rm ann} + \sum_{I}(\sigma_I v_{\rm rel})_{\rm BSF}$. The relevant quantity for the freeze-out, as we have shown, is $(\sigma v_{\rm rel})_{\rm eff} =  (\sigma v_{\rm rel})_{\rm ann} + \sum_{I}(\sigma_I v_{\rm rel})_{\rm BSF}  \BR(\text{B}_I \rightarrow \text{SM})  \leq (\sigma v_{\rm rel})_{\rm total} $. The equality saturates only at zero temperature, otherwise the inequality holds, due to the fraction of bound states broken by ambient plasma quanta.

In Ref.~\cite{GriestKamionkowski}, the total reaction cross section is approximated as $(\sigma v_{\rm rel})_{\rm total}  \approx (\sigma v_{\rm rel})_{\rm ann}$ and taken for the freeze-out computation, not considering the bound-state effects. The scaling, with the inverse velocity of this cross section, is however only possible in the presence of light mediators, which unavoidably lead to bound state formation~\cite{1703.00478}. Thus, in any perturbative physical system, saturating the unitarity bound on $(\sigma v_{\rm rel})_{\rm total}^J$, the inequality 
$(\sigma v_{\rm rel})_{\rm eff} \leq (\sigma v_{\rm rel})_{\rm total} $ leads to a lower maximally attainable DM mass than expected from considering only annihilation. This is one of the main findings of our paper and will be made quantitative in the coming sections. 

The second case considered in Ref.~\cite{GriestKamionkowski} is the annihilation of extended DM objects. Then the total cross section becomes geometrical as a result of the sum over all partial waves allowed in the annihilation process,
\begin{align}
\label{eq:uniboundextended}
(\sigma)_{\rm total} < \sum_{\ell=0}^{\ell_\text{class}} \frac{4\pi (2\,\ell+1) }{p_{\rm DM}^2}  \approx 4 \pi R_{\rm DM}^2 \,,
\end{align}
since by momentum conservation $\ell_\text{class} \approx p_{\rm DM} \,R$. Note that in the case of the geometrical interaction the cross section $\sigma$ is constant and the quantity relevant for the freeze-out scales as $\sigma v_{\rm rel} \propto v_{\rm rel}$. In principle, the required freeze-out cross section now sets a lower bound on the size of the dark object. This is, however, a zero-temperature computation and does not take into account that some angular momentum eigenstates will lose energy less efficiently and be broken by ambient plasma particles prior to annihilation. The situation is then similar to the point-like particle case since higher partial waves contribute to the total cross section at zero temperature, they will not efficiently contribute to annihilation in the hot plasma and thus change the annihilation efficiency. In Sec.~\ref{sec:extended}, we demonstrate how this finite-temperature effect severely alters the obtained bounds.

%%%%%%%%%%%%%%%%%%%%%%%%%%%%%%%%%%%%%%%%%%%%%%%%%%%%%%%%%%%%%%%%%%%%%
%                     Unitarity Bound in Perturbative Models
%%%%%%%%%%%%%%%%%%%%%%%%%%%%%%%%%%%%%%%%%%%%%%%%%%%%%%%%%%%%%%%%%%%%%
%%%%%%%%%%%%%%%%%%%%%%%%%%%%%%%%%%%%%%%%%%%%%%%%%%%%%%%%%%%%%%%%%%%%%
%                    Model
%%%%%%%%%%%%%%%%%%%%%%%%%%%%%%%%%%%%%%%%%%%%%%%%%%%%%%%%%%%%%%%%%%%%%

\section{Unitarity for Pointlike  WIMPs} \label{sec:model}

In this section, we describe the non-perturbative effects in the simplest model of weakly interacting DM. The basic assumption is that DM interacts with SM particles and that, in the case of heavy DM, those particles induce a long-range force if, at the freeze-out temperature, we have $\MDM > M_{\rm SM}/v_{\rm rel} $ and $\MDM > M_{\rm SM}/\alpha$, where $M_{\rm SM}$ is the mass of a weak scale particle and $\alpha$ the strength of the interaction. Explicit examples have been studied in Refs.~\cite{HiggsMediatedBS, CosmoBS}.

To parametrize the explicit model, we use $SU(2)$ symmetry as the guiding principle.  We consider a particle in the $SU(2)_{\rm L}$ representation $R$.  To have a neutral component, the hypercharge has to be $0$ for odd $R$ and $1/2$ for even $R$. For the heavy freezeout computation, the hypercharge gives just a subleading effect and $R \geq 2$ will be treated as a continuous parameter.  For concreteness, we consider real representations of $SU(2)$ and interpolate in between. Furthermore, DM is assumed to be a Majorana fermion, such that the effective number of degrees of freedom $g_\chi = 2 R$. This model, even though specific in terms of $SU(2)_L$ quantum numbers, serves as an excellent template to parametrize a generic model which has a coupling to weak gauge bosons and we expect the general statements to hold universally. The tree level annihilation cross section is
\begin{align}
\frac{\sigma v_{\rm ann}}{\sigma_0 } = \frac{2 R^4 + 17 R^2 -19}{32 R}, 
\end{align}
with $\sigma_0 = \pi \alpha_2^2/\MDM^2$~\cite{MinimalDM}.

\subsection{Sommerfeld Enhancement}

The annihilation rate is enhanced by the Sommerfeld effect, which decomposes into isospin channels as
\begin{align}
\frac{\sigma_{\rm S,ann} }{\sigma_{\rm ann}} = \sum_I f_I \,S_{\rm ann} (\lambda_I)\, \approx  \sum_{I =1,3} f_I \,S_{\rm ann} (\lambda_I)\,,
\end{align}
where the weight factor $f_I$ can be found by explicit computation and is listed in Table \ref{Tab:fSommerfeld}, the $S_{\rm ann}(\lambda)$ are the Sommerfeld factors; see Ref.~\cite{Feng:2010zp} for their explicit form. The $\lambda_I$ is the effective coupling of each channel and controls the interaction strength, this coupling can be expressed in terms of quadratic Casimir operators, which leads to $\lambda_I = 1/2 [2C_2(R) -C_2(I)]$. For any $R$ the combination contains $R \otimes \bar{R} \supset 1 \oplus 3  \oplus ... 2R-1$. Since among these representations the singlet and adjoint channels have the smallest quadratic Casimirs the corresponding channels are the most attractive ones, explicitly $\lambda_1 = (R^2-1)/4$ and $\lambda_3 = (R^2-5)/4$ and, sub-dominantly (for $R \geq 3$), $\lambda_5 = (R^2-12)/4$.  The larger representations are less attractive, or even repulsive with an effective Sommerfeld suppression factor, similar to Ref.~\cite{1611.04606}. We point out that since the computation is performed in the $SU(2)$ multiplet formalism, we stop our integration of the Boltzmann equations at $z \approx 10^4$, because at lower temperatures mass-splitting effects would become relevant~\cite{MinimalDM, 1411.0752}. However, for large DM masses this is a very good approximation and errors remain below the percent level.
\begin{table}[!htbp]
 $$ \begin{array}{cc|ccccccc} \nonumber
\hbox{Reps. \textbf{R}} &  & \bf{2}  & \bf{3} &  \bf{5} & \bf{7} & \bf{9} & \bf{11} & \bf{13}\\ \hline
\hbox{Channels} &  &  &   & \\ \hline
  Singlet    & f_1 & 0.13 & 0.14 & 0.23 & 0.27 & 0.29 & 0.30 & 0.31 \\ \hline
 Triplet    & f_3  & 0.16 & 0.18 & 0.41 & 0.51 & 0.57 & 0.60 & 0.62 \\ \hline
\end{array}
$$
\caption{\label{Tab:fSommerfeld}   Relative weight factors of the Sommerfeld effect for the leading channels. }
\end{table}

\subsection{Bound-State Effects}

For the estimate of the bound-state formation effect, it is sufficient to take into account the bound states in the 1 (singlet) and the 3 (adjoint) channel, as they correspond to the most attractive channels. As discussed in Ref.~\cite{CosmoBS}, the wave function symmetry leads to selection rules, such that the singlet bound state has spin-$0$ for $\ell$ even and spin-$1$ for $\ell$ odd and the adjoint bound state has spin-$0$ for $\ell$ odd and spin-$1$ for $\ell$ even.   In the $SU(2)$-symmetric limit, the singlet bound state can only be formed from an adjoint initial configuration, but the adjoint state can be formed from the singlet initial configuration and, if $R\geq3$, from the $5$ configuration.

\subsubsection{Formation Rates}

The total formation cross section can be written in a similar way to the Coulomb cross section, but taking into account different potentials for the initial and final states and a factored-out non-abelian structure,
\begin{align}
(\sigma v_{\rm rel})_{\ell}^{n} = (\sigma v_{\rm rel})_{\ell}^{n}(\lambda_i, \lambda_f) \times \left| C_J + \gamma_{\ell}^n C_\tau \right|^2\,.
\end{align}
For example, in the non-relativistic limit, the capture cross section to the ground state can be written as 
\begin{align}
\frac{(\sigma v_{\rm rel})_{\rm BSF}^{1,0}}{ \left( \frac{\sigma_0  (2S+1) 2^{11} \pi}{3 \, g_\chi^2} \right) } \approx \left| C_J + \frac{1}{\lambda_f} C_\tau \right|^2 \times \frac{\lambda_i^3 \alpha_2}{\lambda_f v_{\rm rel}} e^{-4 \lambda_i/\lambda_f}\,,
\end{align}
where $S$ is the spin of the bound state and $C_J$ and $C_\tau$  are group theory factors for the abelian and non-abelian emission, respectively. The full expressions, including the excited states, can be found in Ref.~\cite{CosmoBS}. 

The group-theory factors depend on the transition and have in general a complex structure, see Refs.~\cite{CosmoBS,1805.01200} for the full expressions. In the cases of the transitions we are interested in, however, they can be rewritten in a general simple way:
\begin{align}
 & C_J(1 \rightarrow 3) =C_J(3 \rightarrow 1) = \sqrt{\frac{R^2-1}{4}} \, , \\ 
 & C_\tau(1 \rightarrow 3) = - C_\tau(3 \rightarrow 1)= -\sqrt{\frac{R^2-1}{4}} \, , \\
 & C_J(3 \rightarrow 5) =C_J(5 \rightarrow 3) = \sqrt{\frac{R^2-4}{2}}  \,, \\ 
 & C_\tau(5 \rightarrow 3) = - C_\tau(3 \rightarrow 5)= \sqrt{2\left(R^2-1\right)} \,.
\end{align}
With these factors and the $\lambda_I$  for $I =\lbrace 1 ,2,3 \rbrace$, the cross sections can be readily computed. Note that capture to the adjoint state consists of two transitions from the singlet and  (if $R \geq 3$) the $5$ state, which makes it relevant for larger representations. As discussed in Ref.~\cite{CosmoBS}, the effects of massive mediators can be taken into account by a kinematic factor, which will suppress bound-state formation, once the binding energy is not sufficient to emit the gauge boson.

A further aspect to take into account is the formation of excited bound states, which can be taken into account by means of Kramer's formula; see Ref.~\cite{Kramer}. The resummation leads to a logarithmic enhancement of the capture rate with $\approx \left( 1 + \kappa \ln{\left( \alpha_{\rm eff}/2 v_{\rm rel} \right)}\right)$.

\subsubsection{Annihilation and Decay Rates}

From Fig.~\ref{fig:FreezeOutBS}, it becomes clear that the annihilation rates are crucial for judging the impact of a bound state on the final DM relic abundance.  As discussed in Ref.~\cite{CosmoBS}, the excited  $\ell >1$ states have strongly suppressed annihilation rates and decay to the ground states in order to annihilate. We work with selection rules for real $SU(2)$ representations, which have $Y=0$, as pseudo-real representations are only viable as admixtures, as discussed in Ref.~\cite{Tytgat}. Given the selection rules for identical fermions, the singlet ground-state bound states will have spin-$0$ and annihilate to two gauge bosons with a rate of 
\begin{align}
\frac{\Gamma_{1\, S=0}^{n\, \ell=0}}{\MDM } = \frac{\alpha_2^5 C_2(R)^3 T_R^2 d_{\text{adj}}}{2 \eta  n^3 d_R} = \frac{\alpha _2^5  R \left(R^2-1\right)^5}{3  \, \eta \,2^{11}\,n^3}\,,
\end{align}
\noindent where $T_R$ is the representation index. On the other hand, the selection rules dictate that spin-$1$ states are triplets of $SU(2)$ and their annihilation to gauge bosons is suppressed. They will annihilate democratically into SM fermions with the rate 
\begin{align}
\frac{\Gamma_{3\, S=1}^{n \, \ell=0}}{\MDM} = \frac{\alpha _2^5 \lambda_3^3 n_f T_R T_{\text{SM}}}{3 \, \eta \,  2^2 \, n^3} =\frac{n_f\alpha_2^5 R \left(R^2-1\right) \left(R^2-5\right)^3}{9\, \eta\, 2^{11} \,n^3}\,. 
\end{align}
\noindent The total rate contains the SM fermion multiplicity $n_f = 3(3+1)$, $T_{\rm SM} =1/2$ and $\eta= 2$ for self-conjugrate DM and $\eta= 1$ for non-self-conjugrate DM. 

\begin{figure}[t]
\includegraphics[width=0.45\textwidth]{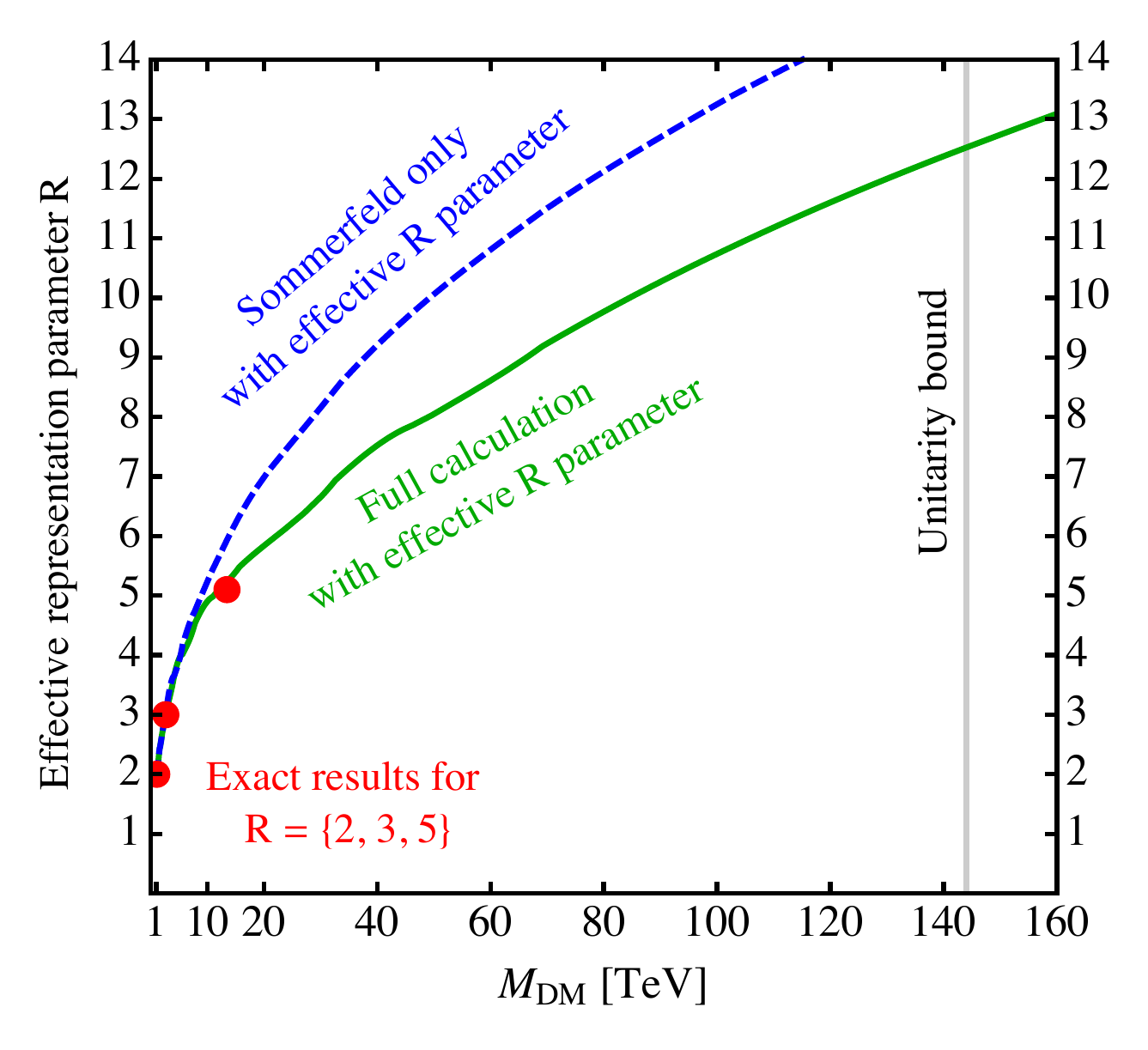}
\caption{\label{fig:ParametericWIMP}  The green solid line indicates the mass and representation at which the relic density for a parametric WIMP is equal to the observed DM density, including bound-state formation.  The blue dashed line shows the same prediction with only Sommerfeld enhancement taken into account.  The largest representation compatible with unitarity and the DM abundance is $R=12$.}
\end{figure}

In Fig.~\ref{fig:ParametericWIMP}, we show the DM mass that leads to the correct DM relic abundance given a representation $R$. We see that our estimates agree well with the full computation for the  $R =\lbrace 2, 3,5 \rbrace$ representation. In addition, we show that for $R \geq 4$, the bound-state formation effects lead to substantial corrections on top of the Sommerfeld corrections.  

\subsubsection{Partial-Wave Contributions}
 
As already discussed in Ref.~\cite{GriestKamionkowski}, the annihilation cross section $(\sigma_{\rm ann} v)$ is dominated by the s-wave annihilation process. We perform analogously the partial wave decomposition of the amplitudes for the bound-state formation cross sections. This is done by obtaining the different $J$ contributions by projecting the transition amplitude on the Legendre polynomials. For example the $(\sigma_{\rm BSF} v)^{n=2, \ell =1 }$ contains a significant $s$-wave fraction, in agreement with~\cite{1611.01394}. By explicit computation, it can be verified that the amplitudes for higher excited states, $(\sigma_{\rm BSF} v)^{n \ell}$, when summed over $\ell$, contain the same s-wave contributions. Those states are the ones relevant for the derivation of Kramer's resummation formula. We find that the s-wave processes contribution to the total cross section is thus given by
\begin{align}
(\sigma v)_{\rm total}^0 =(f_{\rm ann }^0)^2  (\sigma_{\rm ann} v)  + \sum_I  (f_I^0)^2 \, (\sigma_{\rm BSF}^I v)\,.
\end{align}
This cross section will be used to find the value of the coupling strength, controlled by $R$, at which the $s$-wave contribution saturates the unitarity bound. 
\begin{figure}[t]
\includegraphics[width=0.5\textwidth]{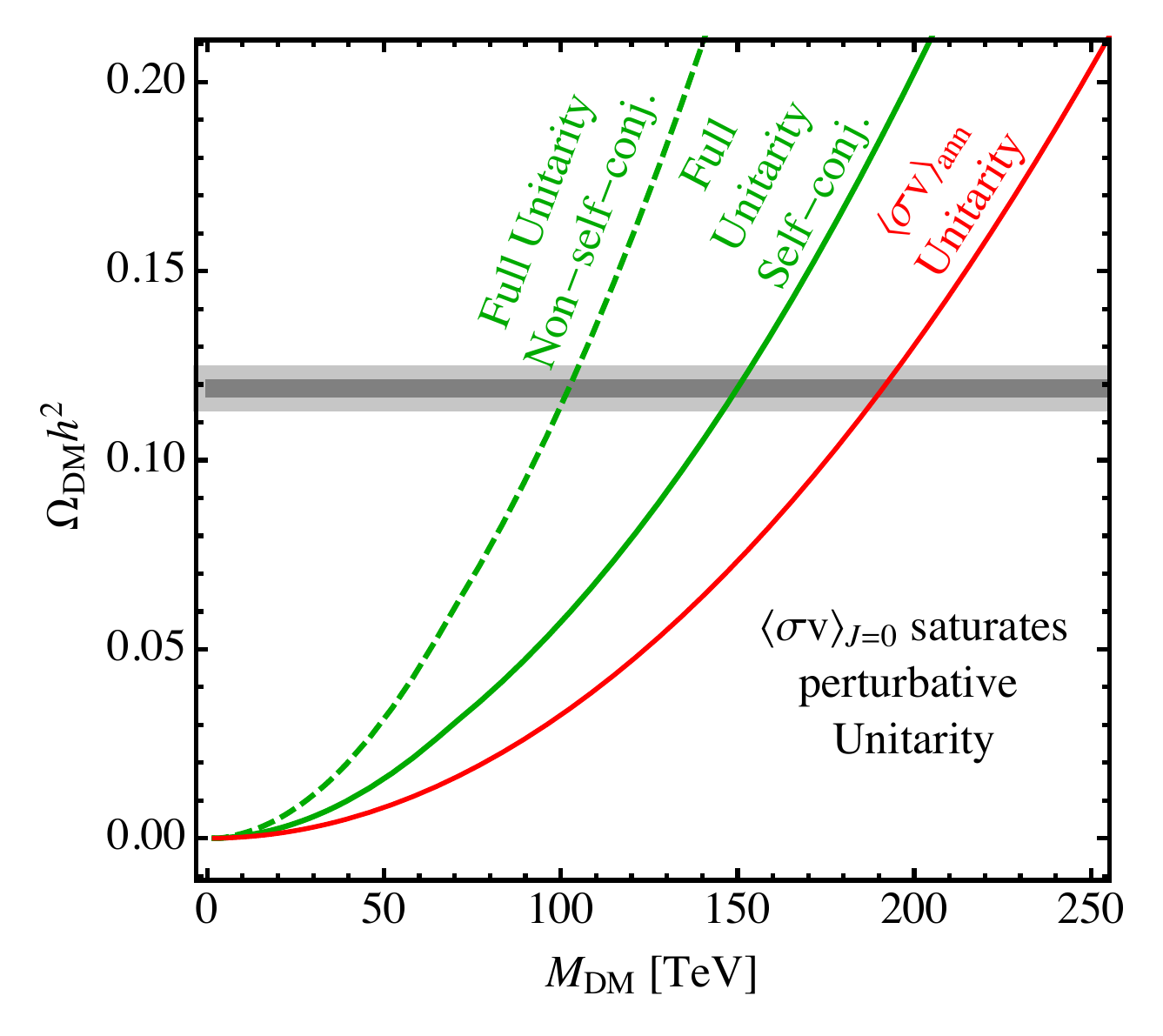}
\caption{\label{fig:RelicD}  The green solid line shows the predicted relic density under the assumption that the s-wave cross section saturates the full unitarity bound, for a self-conjugate DM candidate. The green dashed line is the same for the non-self-conjugate case.  The red line shows this prediction when bound-state formation is neglected. }
\end{figure}

%%%%%%%%%%%%%%%%%%%%%%%%%%%%%%%%%%%%%%%%%%%%%%%%%%%%%%%%%%%%%%%%%%%%%
%                Implications
%%%%%%%%%%%%%%%%%%%%%%%%%%%%%%%%%%%%%%%%%%%%%%%%%%%%%%%%%%%%%%%%%%%%%

\subsection{ Unitarity Bounds for Pointlike WIMPs} \label{sec:implications}

Each value of the $R$ parameter corresponds to a DM model where the lightest new particle can be a superposition of fermions but has a dominant contribution, which transforms in the $R$ representation of $SU(2)_L$. Therefore, the only free parameter for such a model is the mass of the DM candidate, and we determine it to satisfy the observed relic-density value.

%%%%%%%%%%%%%%%%%%%%%%%%%%%%%%%%%%%%%%%%%%%%%%%%%%%%%%%%%%%%%%%%%%%%%
%                    Mass Range reduced!!!
%%%%%%%%%%%%%%%%%%%%%%%%%%%%%%%%%%%%%%%%%%%%%%%%%%%%%%%%%%%%%%%%%%%%%

In Fig. ~\ref{fig:RelicD}, we show the relic abundance as a function of the DM mass given that its cross section saturates the unitarity bound. The computation is performed in the following way: we demand that $(\sigma v)_{\rm total}^0 < (\sigma v)_{\rm max}^0$, which is set by partial wave unitarity, Eq.~(\ref{eq:unibound}), and extract for each given DM mass a value for  $R_{\rm max}$. With this maximal coupling and the DM mass, we compute the relic abundance using now the full bound-state formation cross sections, not limiting them to the $s$-wave processes.  The allowed DM mass range is found to be $\MDM \leq 144 \, \rm TeV$ ($\MDM \leq  96 \, \rm TeV$ for a non-self-conjugate particle). We have to stress that the obtained mass bound is subject to significant uncertainty since the effective coupling values are large. Thus higher-order processes, such as multiple gauge boson emission, are likely to become relevant. The DM mass bound is an important result of this paper and applies to asymmetric DM as well. Note that without taking into account the bound-state formation, which helps to saturate the unitarity bound, but does not fully contribute to the annihilation cross section, the allowed mass range would be significantly higher, $\MDM \leq 200 \, \rm TeV$. 

This allows an interesting argument for new physics at the 100-TeV scale.  As a historical example, the breakdown of perturbative unitarity in the gauge boson scattering cross section near 1 TeV indicated the existence of the light SM Higgs boson \cite{HiggsUnitarity}.  Here, the upper bound on the DM annihilation cross section is a consequence of probability conservation only.  In the case of a pointlike WIMP, this bound induces an upper bound on the WIMP mass and thus renders the WIMP parameter space finite. Furthermore, as we have demonstrated, the scale of new physics, containing a least the WIMP particle is below 100 TeV. This mass scale is low enough to be explored by future collider experiments. This conclusion is reached without appealing to naturalness or other aesthetic considerations. 

\begin{figure*}[tb]
\centering
\subfloat[The allowed cross section range in the early universe. The bound-state formation is suppressed below $5$ TeV since the bound-state binding energy is too low to emit $W$ and $Z$ bosons and the process is kinematically blocked. Thus the $J=0$ unitarity bound coincides with the tree-level result.  Note that the unitarity limit on the cross section is inversely proportional to the DM velocity and is thus more stringent in the early universe. The total DM annihilation cross section, denoted by the blue line, is also lower in the early universe.]
{\includegraphics[width=.49\textwidth]{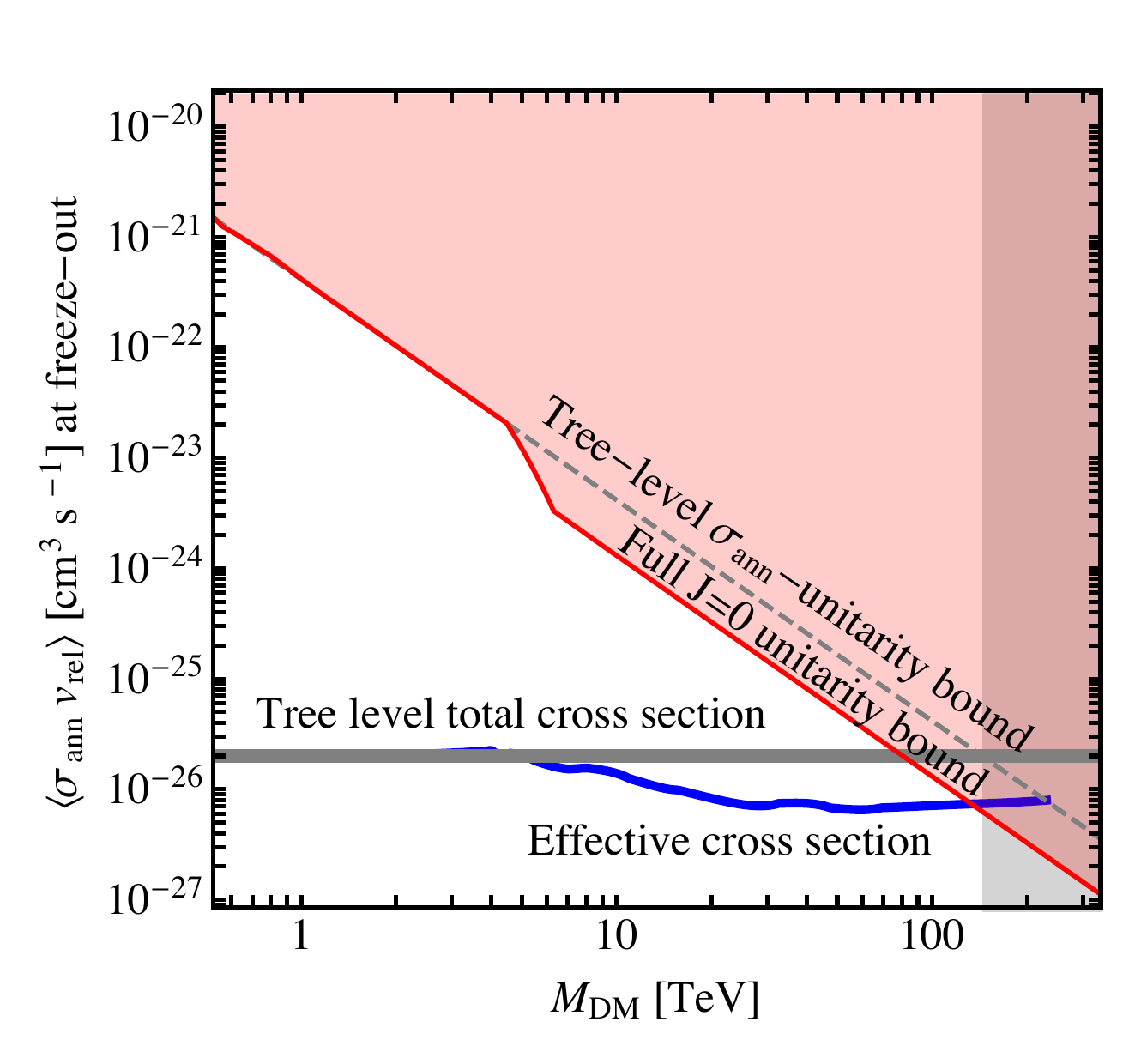}}
\hfill
\subfloat[
The implications of the full unitarity bound on the DM cross section today. Superposed are limits from the Fermi-LAT dwarf spheroidal observations from Ref.~\cite{FermiLAT}. The blue line shows the direct annihilation of DM and $W$-bosons, including the decay of spin-$0$  bound states. The $\gamma$-line cross section indicated here by the orange, dashed line arises from the bound-state formation process and the photons are emitted at $E_\gamma \approx E_\text{B}$. The purple and green, dashed lines show the annihilation cross section to light fermionic final states.]
{\includegraphics[width=.49\textwidth]{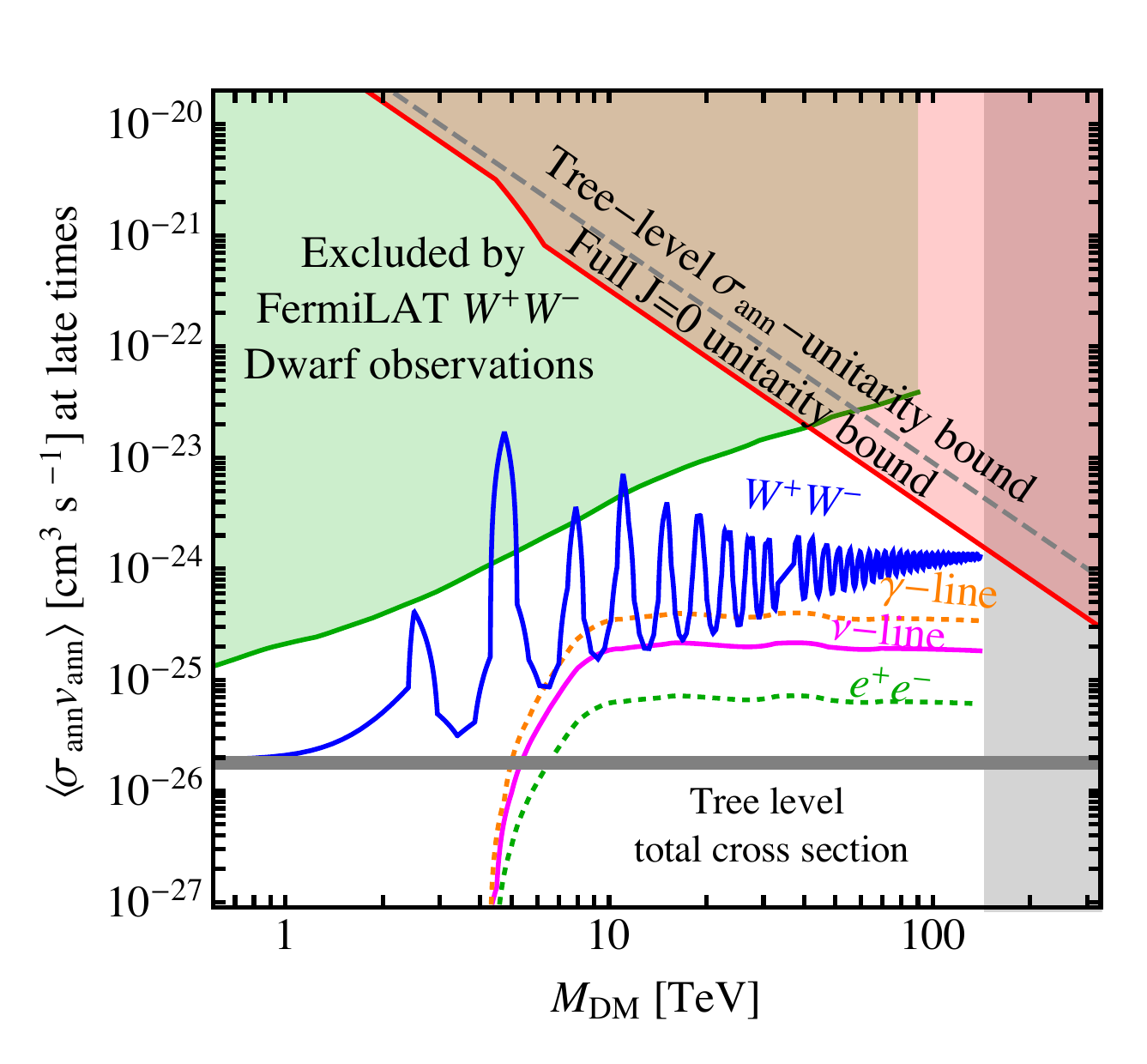}}
\caption{\label{fig:ExclussionsLateEarly}  Comparison of the early universe and late-time annihilation cross sections allowed by unitarity and experimental constraints. Since the cross section contains the $1/v_{\rm rel}$ factor from the Sommerfeld and bound-state formation effects, the unitarity bounds on the maximally allowed cross section strongly differ between the early universe in (a)  and late time annihilation in (b).  Today, the annihilation cross section can be significantly enhanced.}
\end{figure*}

%%%%%%%%%%%%%%%%%%%%%%%%%%%%%%%%%%%%%%%%%%%%%%%%%%%%%%%%%%%%%%%%%%%%%
%                    U(1)
%%%%%%%%%%%%%%%%%%%%%%%%%%%%%%%%%%%%%%%%%%%%%%%%%%%%%%%%%%%%%%%%%%%%%

\subsection{The Abelian WIMP} \label{sec:previous}

Previous papers investigated the freezeout of fermionic DM interacting via a dark Abelian force, also in the case of asymmetric self-interacting DM~\cite{1712.07489}.  While Ref.~\cite{Petraki} concluded that taking into account bound-state formation reduced the unitarity limit on the DM mass to $139$ TeV, a later study suggested that this was incorrect~\cite{1611.01394}.
 
Here, we consider the impact on freezeout of bound states with spin-1 or spin-0 with $n=1$. It is argued that at leading order the selection rule $\Delta \ell =1$ leads to $p$-wave dominance of the bound-state formation in the ground state. Therefore, the bound-state formation does not contribute to the saturation of the $s$-wave unitarity bound and the maximal dark gauge coupling allowed by unitarity remains $\alpha_D < 0.86$, which is determined only from the Sommerfeld enhanced annihilation cross section. This is the reason, why in~\cite{1611.01394} no DM mass reduction was found. 
 
The argument based on the leading order selection rule is, however, problematic given the size of the gauge coupling involved. We therefore investigate the effect of two-photon emission adopting the formalism of Ref.~\cite{Bhanot:1979vb}. The $\Delta \ell =0 $ contribution to the bound-state formation in the ground state is $s$-wave dominated and given by
\begin{align}
 & (\sigma_{n=1,\ell =0}^S  v_{\rm rel} )= \frac{3 \,2^4\left(2 S+1\right) \alpha_D^2 }{\pi\, g_\chi^2}   \int_{0}^{\Delta E}dk\,k^3 \left(\Delta E-k\right)^3   \nonumber \\
&\times  \sum_{n=2}^\infty \left| \langle \phi_{\ell=0}|r |  \psi_{n,1}  \rangle   \langle \psi_{n,1}|r |  \psi_{1,0}  \rangle  \right|^2 \\ \nonumber
& \left\{\frac{1}{-E_{1,0}+E_{n,1}-k}+\frac{1}{-E_{1,0}+E_{n,1}-(\Delta E-k)}\right\}^2 \\ \nonumber
& \approx \frac{\tilde{\kappa}  \,\left(2 S+1\right) \bar{\sigma}_{0} \, \alpha_D^3}{g_\chi^2} \frac{\alpha_D}{v_{\rm rel}}\,,
\end{align} 
where $\phi$ is the radial part of the free two-particle state wave-function, $\psi$ are the radial parts of bound-state wave-functions, the numerical factor $\tilde{\kappa} \approx 20$ and $\bar{\sigma}_0 = \pi \alpha^2_D/\MDM^2$. Taking this contribution to the ground-state formation into account reduces the $s$-wave unitarity bound on the dark gauge coupling to $\alpha_D < 0.6$ and leads furthermore to a reduced maximally attainable DM mass of the order of $\MDM < 150 \,\rm TeV$. (Compared to the bound of $\MDM < 200 \,\rm TeV$ if only the Sommerfeld effect was considered).  This finding is in agreement with the expectations that we elaborated on earlier. Because bound-state formation contributes to the total inelastic cross section, it reduces the maximum viable coupling strength, but because it is not always efficient in reducing the DM abundance, the bound on the maximal DM mass is decreased. This example shows at the same time that close to the unitarity bound, the large coupling strength makes higher-order processes relevant and therefore the computed cross sections are expected to have a substantial theoretical uncertainty.

\begin{figure*}[tb]
\centering
\subfloat[The low-energy gamma-ray spectrum, which shows the gamma ray lines from capture photons on top of the continuum fragmentation spectrum computed with~\cite{1012.4515}]
{\includegraphics[width=.48\textwidth]{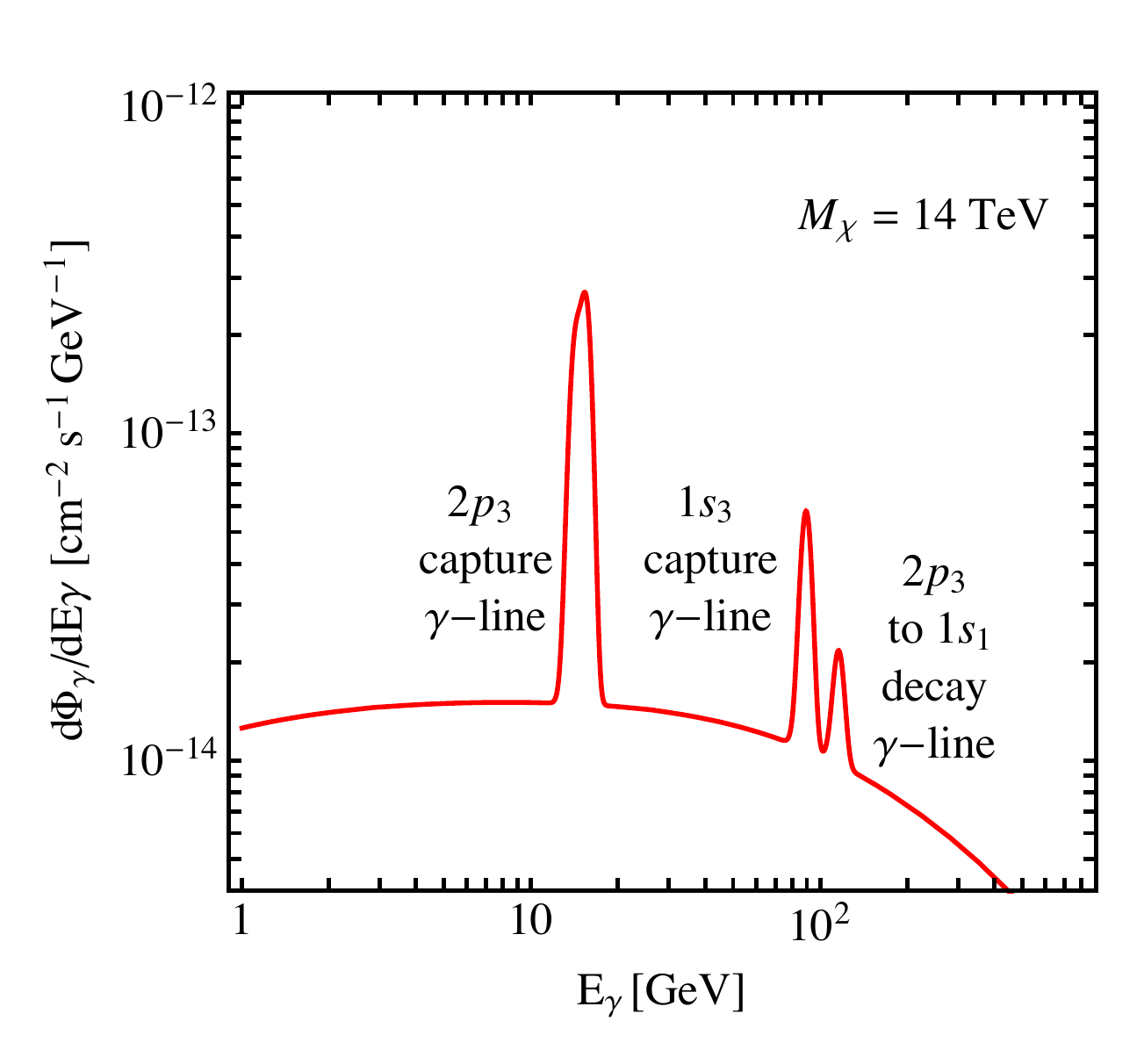}}
\hfill
\subfloat[
The high-energy gamma-ray spectrum with direct annihilation to gamma ray lines and fragmentation photon spectra.]
{\includegraphics[width=.49\textwidth]{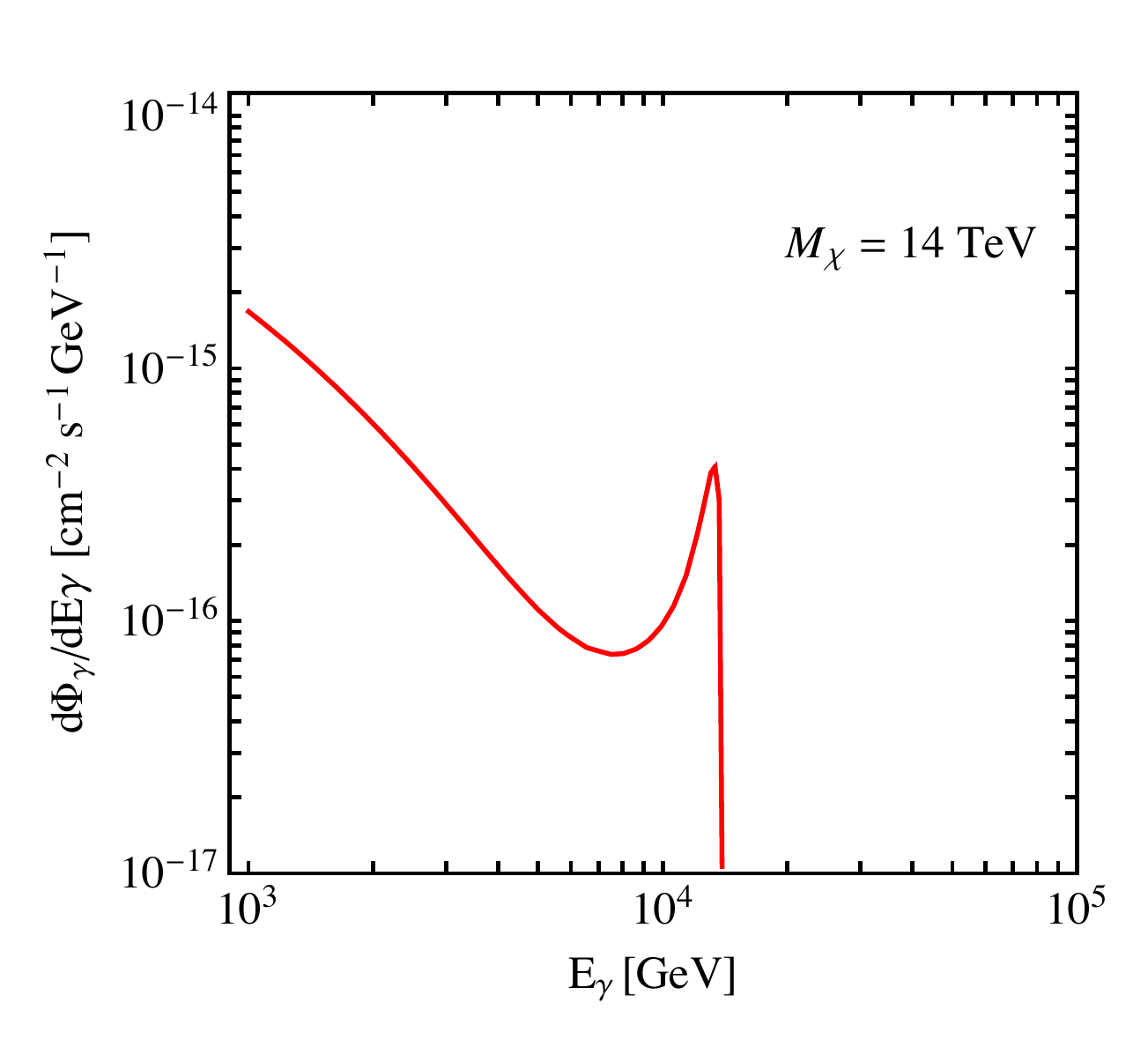}}
\caption{\label{fig:Spectra}  Gamma-ray spectra from a $SU(2)$ DM candidate with $R=5$ and a mass of $14$ TeV.  The energy resolution is taken to be $10\%$, corresponding to the Fermi LAT experiment. For the flux a J-factor of the galactic center region $R_3$ of $13.9\cdot10^{22}\, \text{GeV}^2 \text{cm}^{-5}$ was assumed. The large separation of monochromatic signals in their energy makes experimental searches for them highly complementary.}
\end{figure*}

%%%%%%%%%%%%%%%%%%%%%%%%%%%%%%%%%%%%%%%%%%%%%%%%%%%%%%%%%%%%%%%%%%%%%
%                    Bounds 
%%%%%%%%%%%%%%%%%%%%%%%%%%%%%%%%%%%%%%%%%%%%%%%%%%%%%%%%%%%%%%%%%%%%%

\subsection{Observable Signatures} \label{sec:bounds}

The effective WIMP framework makes concrete predictions for various channels and connects different observables. We show that several unexpected new signatures arise.
This predictive power is the result of $SU(2)_L$ gauge symmetry, which we used to construct the effective WIMP model. 

A new exciting possibility for indirect detection is the observation of gamma rays emitted in the formation of the dark-matter bound states. Those capture gamma rays are nearly monochromatic and carry away the binding energy.

A few generic statements can be made about this new process. It is generally true that the neutral component of the weak multiplet, which is the lightest particle in the spectrum and the natural DM candidate, has no triplet component in the wave-function.  Therefore, no bound state can be formed in the singlet configuration at zero temperature. 

The largest formation cross sections will thus be attributed to the adjoint, i.e., the triplet configuration. The most probable main quantum numbers will be the ground state, with $n=1$, i.e., the $1s_3$ state and the lowest excited state, with $n=2$ and $p=1$, i.e., the $2p_3$ state. The production cross section for the latter is sizable, since it is produced from the s-wave configuration and is not suppressed at low velocities. Furthermore, each excited state will decay to the ground state before annihilation, in particular there will be the $2p_3$ to $1s_1$  transition. We can predict, for each representation $R$, the relative position of the dominant gamma lines in the Coulomb limit 
\begin{align}
& \frac{E^\gamma_{1s_3}}{ \MDM} = \frac{\lambda_3^2 \alpha_2^2}{4}  =\frac{\alpha_2^2\,(R^2-5)^2}{64}  \, , \nonumber \\
& \frac{E^\gamma_{2p_3}}{ \MDM} = \frac{\lambda_3^2 \alpha_2^2}{16}  =\frac{\alpha_2^2\,(R^2-5)^2 }{256} \, ,\nonumber \\  
& \frac{E^\gamma_{2p_3 -1s_1}}{ \MDM} = \frac{\alpha_2^2}{256} \left( 3 R^4 + 2 R^2 -21 \right) \,.
\end{align}

In Fig.~\ref{fig:ExclussionsLateEarly} (a), the annihilation cross section at freeze out in the early universe is shown. The cross section, in the presence of long-range forces scales as $1/v_{\rm rel.}$. Since the average velocities at freeze-out are of order $v_{\rm rel.} \sim \mathcal{O}(0.3)$, the cross section is close to the well known value of $2.2 \times 10^{-26} \, \text{cm}^3 \, \text{s}^{-1}$. In contrast, the late time annihilation is significantly enhanced, since the average DM velocity in galaxies today is $v_{\rm rel.} \sim \mathcal{O}(10^{-3})$. This is indicated in Fig. \ref{fig:ExclussionsLateEarly} (b) for all allowed final states. 

In Fig. \ref{fig:ExclussionsLateEarly} (b) ($\gamma-$line), the capture gamma line cross sections are shown as a function of DM mass.  Measuring the ratio of those three line energies provides enough spectroscopic data to determine the $R$ parameter and thus, the DM gauge charge. This seems to be a unique possibility, in which indirect detection experiments can probe not only the magnitude of the annihilation cross section to test the freeze-out hypothesis but also directly measure the DM connection to the SM. Since given the large DM mass the capture photons emitted will be also hard gamma rays, resummation techniques can become important for the precise spectrum prediction~\cite{1712.07656, 1808.08956}. Because at late times DM entirely consists of the neutral $\chi_0$ component, we have to project out its $SU(2)$ multiplet components to compute the late time annihilation signals. In our analysis this is done by use of the Clebsch-Gordan coefficients. This procedure was already used in Ref.~\cite{CosmoBS} and a direct comparison of 5-plet spectra to previous exact results~\cite{0706.4071, 1507.05519,1608.00786} shows that it is a good estimator for the order of magnitude signal strength.

In Fig. \ref{fig:Spectra} panel (a), we show the capture photon spectra for a $14$ TeV DM candidate with $R=5$ on top of the gamma-ray continuum.  The figure demonstrates that given a $10\%$ energy resolution, the lines will be visible. Panel (b) shows the high energy line signal  from direct annihilation, which is a particularly good target for the H.E.S.S observatory \cite{1810.00995}. We point out that because the signals are strongly separated in their respective energy range, the experimental searches for them will be complementary.

\begin{figure}[tb]
\centering
\includegraphics[width=.45\textwidth]{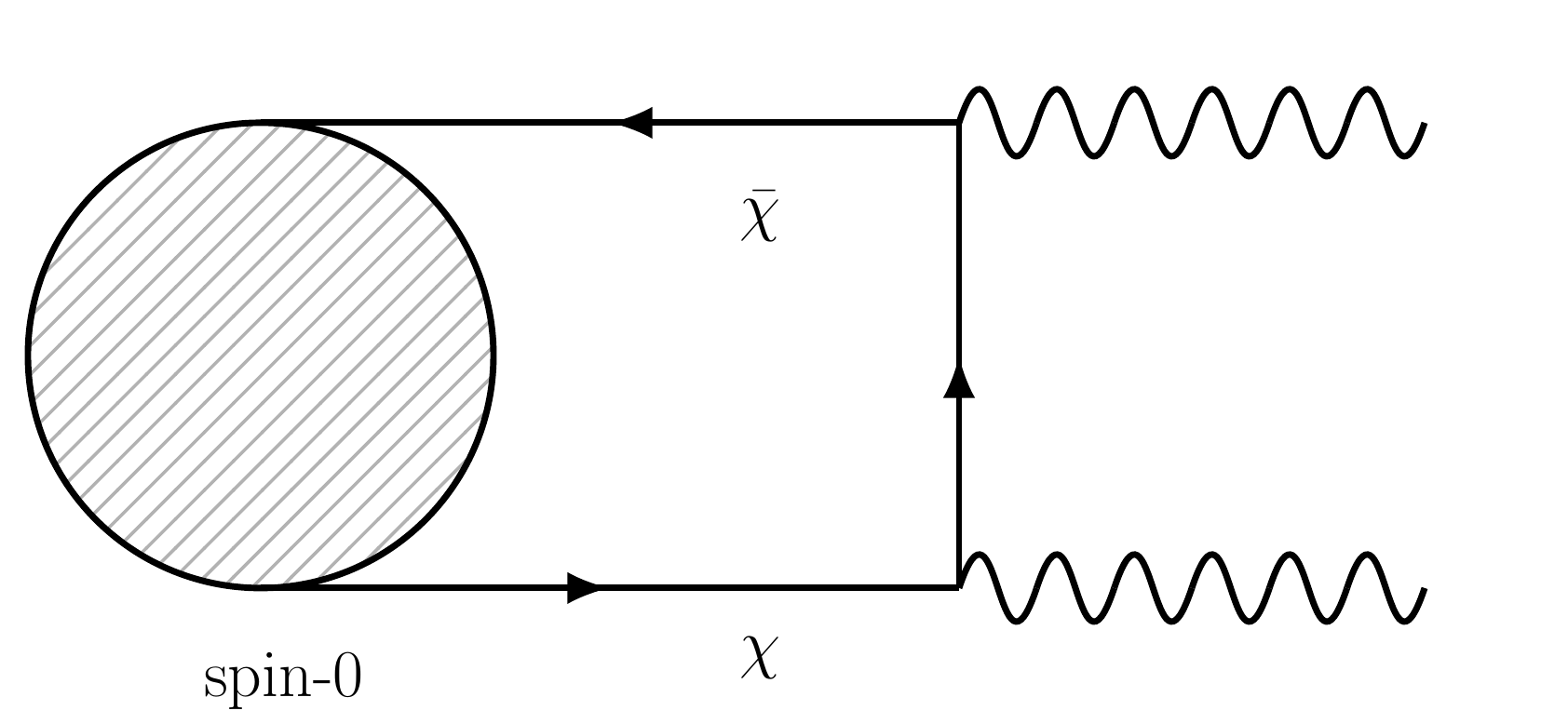}
\caption{\label{fig:FeynmanSpin0} Spin-zero bound-state annihilation to two gauge bosons ($\gamma$, Z, W), where the last requires that the internal line be a charged state of the $\chi$ multiplet.}
\end{figure}

In Fig.  \ref{fig:FeynmanSpin0}, the spin-0 bound state annihilation in shown, which contributes to the same final states as the dominant direct WIMP annihilation channel, and thus enhances the expected signals.  
\begin{figure}[tb]
\centering
\includegraphics[width=.45\textwidth]{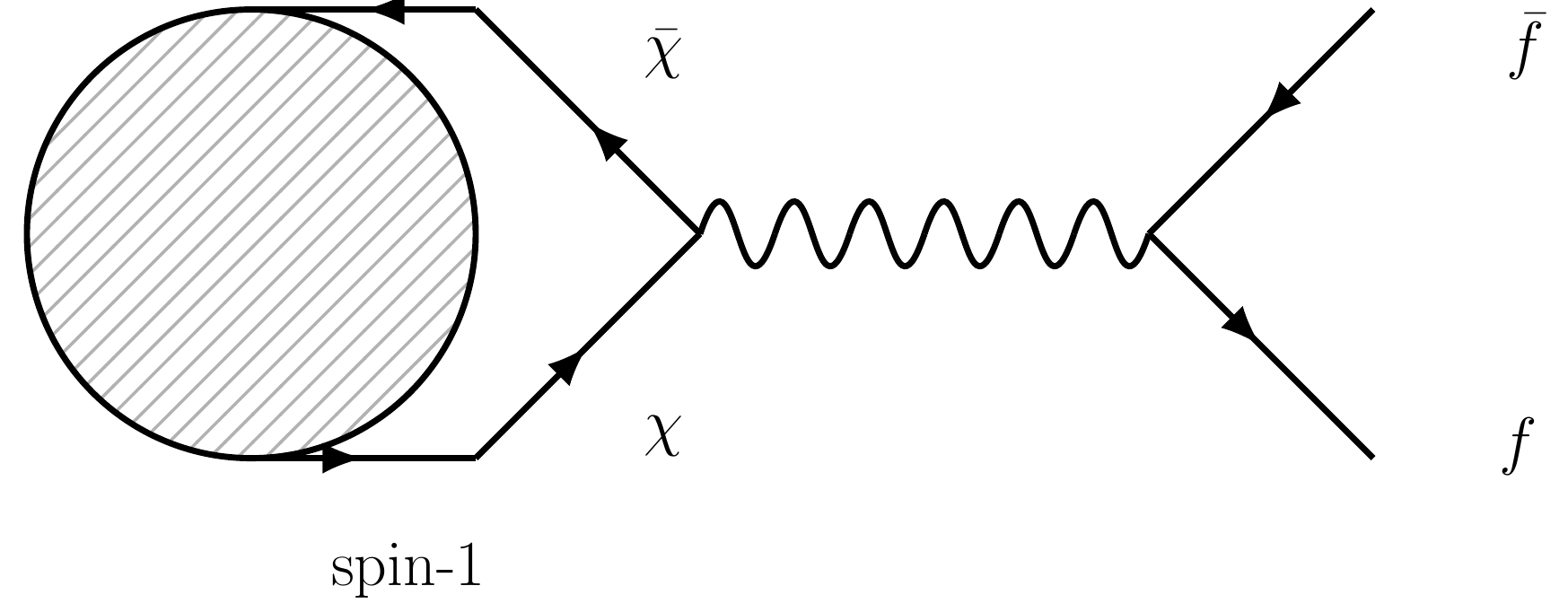}
\caption{\label{fig:FeynmanSpin1} Spin-one bound-state annihilation to two fermions via a gauge boson ($\gamma$, Z).  $SU(2)_L$ symmetry dictates a democratic distribution of fermionic final states.}
\end{figure}

In Fig.~\ref{fig:FeynmanSpin1}, we show another new avenue, which the bound states open up: reactions with light final states. In many models, annihilations to neutrinos or electrons are suppressed due to the smallness of their mass (chirality suppression, e.g., Ref.~\cite{LeptophillicDM}) or due to the smallness of their Yukawa couplings to the Higgs, e.g., Ref.~\cite{DirectVSIndirect}.  
On the contrary, weak WIMP bound states with spin-1 cannot decay to two gauge bosons, and have suppressed annihilation channels to three gauge bosons. The dominant annihilation channel for those states will thus have two SM fermions in the final state.  The $SU(2)_L$ invariance dictates that $25\%$ of those will be leptons and that half of those leptons will be neutrinos. In contrast to spin-1 bound-state annihilation, the direct WIMP annihilation to leptonic final states is subdominant for large representations~\cite{MinimalDM}. Given the current low energy resolution of neutrino observatories, the fact that the produced neutrinos are monochromatic is less relevant than the overall enhancement of the high-energy neutrino flux.

In Fig.~\ref{fig:ExclussionsLateEarly} (b), we show the predicted cross sections for those nearly monochromatic, high-energy neutrinos. These deserve particular attention since they will be easier to distinguish from power-law astrophysical backgrounds. An upper bound on the DM annihilation cross section using neutrino signals was already derived in Refs.~\cite{UpperBoundCross, NuBoundsYuksel}. It is intriguing that at DM masses approaching the unitarity bound, non-perturbative effects unavoidably lead to an enhancement, raising the value of the annihilation cross section above the naively expected one. 

Finally, it is possible to predict the spin-independent scattering cross section with nucleons, needed for the direct detection experiments, as discussed in Ref.~\cite{RunningJap}. The basic assumption is that the DM candidate does not couple directly to the Z boson. This is always true for representations with odd $R$, while representations with even $R$ can be viewed as variants of two or more representations, which mutually mix due to Higgs Yukawa interactions. An example is a Higgsino-like neutralino in supersymmetric models. The general framework for such models has been discussed in Ref.~\cite{Tytgat}.  

\begin{figure}[t]
\includegraphics[width=0.45\textwidth]{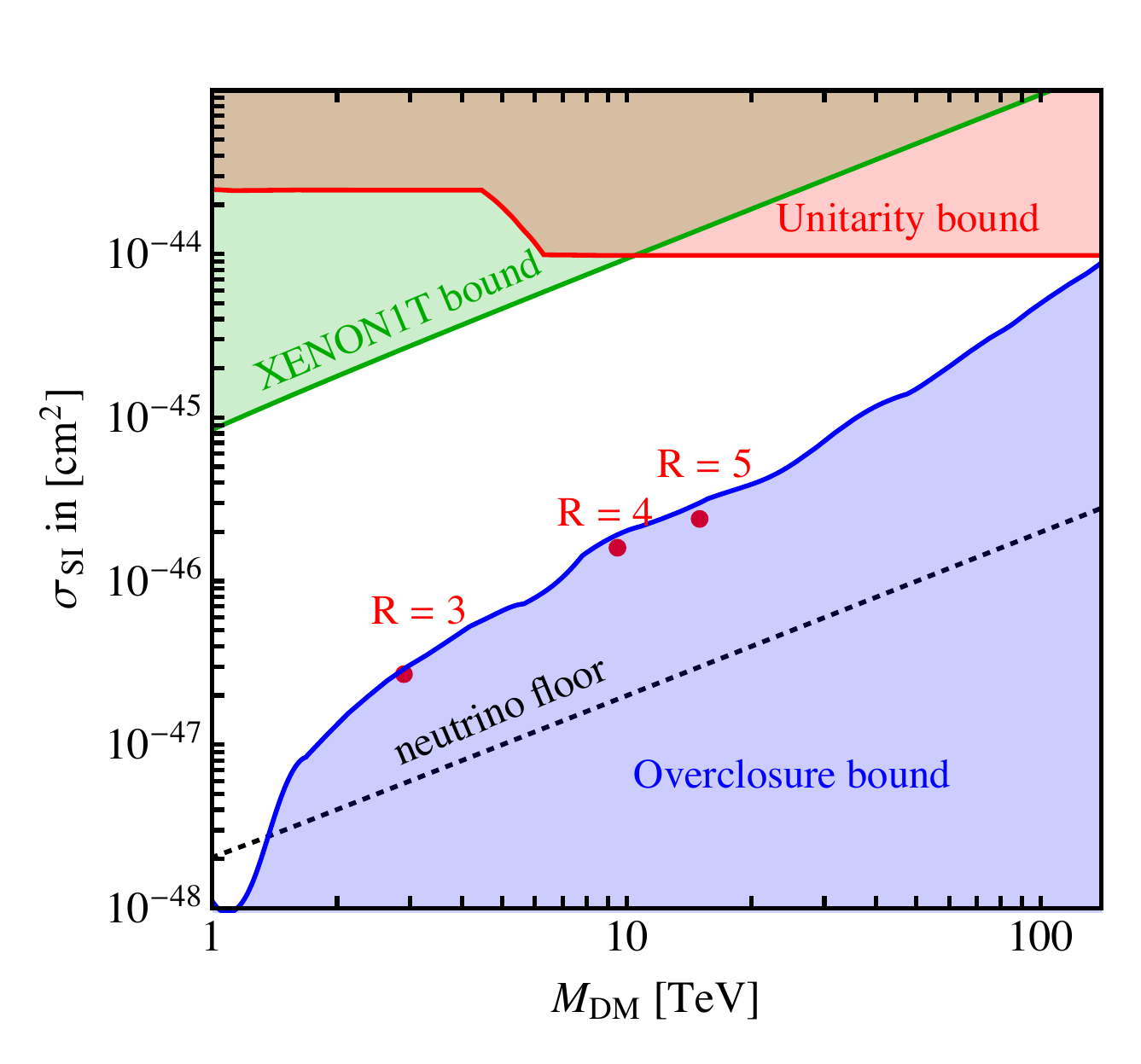}
\caption{\label{fig:DirectDetection}  The predicted spin-independent cross section, under the assumption that DM relic abundance is satisfied. Red dots show the exact results for the $R=\lbrace 3, 4, 5\rbrace$ representations. Superposed are the latest limits of the XENON1T experiment from Ref.~\cite{XENON1T}.}
\end{figure}

In Fig.~\ref{fig:DirectDetection}, we show the predictions for the spin-independent cross section as function of the representation parameter $R$. It is encouraging that almost the entire parameter space of the weakly coupled DM models predicts direct detection cross sections above the neutrino floor. This is an interesting feature of the suggested framework since the available parameter space is finite due to theoretical considerations. It is even more intriguing, since upcoming direct detection experiments, as the DARWIN experiment~\cite{DARWIN}, will cover it entirely. 

%%%%%%%%%%%%%%%%%%%%%%%%%%%%%%%%%%%%%%%%%%%%%%%%%%%%%%%%%%%%%%%%%%%%%
%                    Theoretical consistency 
%%%%%%%%%%%%%%%%%%%%%%%%%%%%%%%%%%%%%%%%%%%%%%%%%%%%%%%%%%%%%%%%%%%%%

\subsection{Theory for Large Representations} \label{sec:safety}

For DM candidates close to the unitarity limit, there is a question of theoretical consistency because, at large coupling constants of large representations, low-lying Landau poles could render the models nonviable; see Ref.~\cite{MinimalDM}. Since in that case, a DM theory needs a low-lying cut off scale,  DM stability might well be compromised. However, recent developments in quantum field theory show that the paradigm of so-called triviality could be challenged. 

As discussed in Ref.~\cite{BubbleSum}, new resummation techniques indicate that it is possible that a theory that exhibits a strong renormalization flow running enters an asymptotically safe fixed point. In particular, it has been demonstrated that in an explicit construction with vector-like fermions, the weak coupling constant has a safe fixed-point. The computation relies on an expansion in the small parameter $1/N_f$ , where $N_f$ is the number of fermion flavors; see Ref.~\cite{Holdom} for the resummation technique.  

A different argument, presented in~\cite{1512.05353}, shows that introducing a millicharge $\epsilon$ instead of the charge neutrality requirement for DM leads to a stable DM candidate without the accidental symmetry assumption. Thus, even in an EFT with sub-planckian Landau pole electroweak multiplets in larger representations are viable DM candidates.

Those considerations open up an entirely new realm of models that have been previously ignored. Concretely, it implies that fermions in $SU(2)_L$ representations larger than the quintuplet are good, minimal DM candidates.  Conversely, the discovery of a DM particle in a large representation would be a smoking-gun signal of an asymptotically safe model.

%%%%%%%%%%%%%%%%%%%%%%%%%%%%%%%%%%%%%%%%%%%%%%%%%%%%%%%%%%%%%%%%%%%%%
%                    Extended objects
%%%%%%%%%%%%%%%%%%%%%%%%%%%%%%%%%%%%%%%%%%%%%%%%%%%%%%%%%%%%%%%%%%%%%

\section{Unitarity Bounds for Extended Objects} \label{sec:extended}

In this section, we show that bound states are also crucial for the unitarity bound for composite DM systems. As discussed in Ref.~\cite{GriestKamionkowski}, the unitarity argument can be also applied to the annihilation of extended composite objects, such as DM atoms, see for example Refs.~\cite{Khlopov:2011tn, Khlopov:2014bia}.  These particles can be seen as WIMPs in the wider sense since they are massive, have initial thermal equilibrium with the SM plasma and a cross section, weak enough such that a significant relic abundance survives asymptotically. The intuitive reason why their cross section is big is that all partial waves up to $l_{\rm class} \approx  p \times R$ contribute, where $R$ is the size of the object and $p$ its momentum. For example, the atomic cross section for hydrogen is large, since the Bohr radius $R \propto 1/ \alpha_{\rm em} m_e$  is controlled by the mass of the electron, which is much smaller than the mass of the entire atom.   

\subsection{The Simplest Dark Atom}

In the case of heavy, multi-TeV DM, the low number density makes hydrogen-like recombination very inefficient~\cite{Tesi}. Thus if heavy DM is an atom, it is formed by a non-abelian confining gauge force. The simplest model for this is an $SU(N)$ gauge theory with a fermion in the adjoint representation. Such a set-up is present in any non-abelian supersymmetric Yang-Mills theory. The minimal example is an $SU(3)$ theory with a fermion $\Q$ in the $8$ representation, as introduced in Refs.~\cite{gluino, gluequark}.  

The fermion can have an explicit vector-like mass. Below this mass scale, the running of the gauge coupling $\alpha$ will lead to a strong coupling regime at a scale $\Lambda$ given by $\Lambda \approx M_\Q \exp{\left(-6 \pi/11 N \alpha(M_\Q) \right)}
$.  
That means that at energies below $\Lambda$ the fermion will be forced to form bound states. 

In the relevant mass regime for DM, the fermion number density is low. Therefore, at the phase transition string breaking will lead to the formation of bound states of adjoint fermions and dark gluons $B = \Q g$. The stability of this state is guaranteed by an accidental global symmetry.   The size of the bound state is set by the confinement scale $R \approx 1/\Lambda$, but its mass is controlled by $M_\Q \gg \Lambda$ and its de-Broglie wavelength is much shorter than its size, as in the case of hydrogen.

\subsection{The Atomic Relic Abundance}

Annihilation happens via $\Q g + \Q g \rightarrow \Q\Q + gg$, where the $G= gg$ is the glueball, the lightest degree of freedom in this theory, and the $\Q\Q$ state can self-annihilate efficiently once it has fallen to the ground state.  Given that $M_\Q \gg \Lambda$, the deeply bound states of $\Q\Q$ can be treated perturbatively, as heavy quarkonium. 

The fermions in the adjoint representation can arrange into a singlet or an adjoint bound state. Higher bound-state representations will be less bound or even repulsive, as is the case for $SU(3)$. Here the two octets can arrange in the di-quark $\Q\Q$ state in the $1 \otimes 8_S \otimes 8_A$ representations, arrangements in higher representations are repulsive. Pure combinatorics suggests that the adjoint configurations are formed with a probability $p_{\rm adj.} = 1/(N^2-1)$ and the singlet with $p_1 = 1/(N^2-1)^2$. Furthermore, the adjoint bound state can lose energy efficiently by radiating a gluon in the adjoint representation, while the singlet has to emit at least two gauge bosons, which is a higher order process.  Therefore, it is sufficient to focus on the reactions of the adjoint configuration. 

Now we use the principles of energy and momentum conservation to derive a reaction cross section bound for the extended object. The unitarity limit on the dark-matter mass will be a direct consequence. 

The bound state is formed in all partial waves allowed by the energy conservation and momentum conservation. The energy balance of the reaction is $2 E_B^{\Q g} + E_{\rm kin}^{\rm CMS} \geq  E_B^{\Q \Q} + M_G $. Assuming that for the  binding energies of the non-perturbative states, the energy of the gluon field in the octet configuration of $\Q\Q$ and the glueball mass $M_G$ are of the order of $\Lambda$, we have the condition for the binding energy between the heavy quarks $E_B^{\Q \Q} <  E_{\rm kin}^{\rm CMS} \approx T < \Lambda$, as we consider the post-confinement regime.

Heavy quarkonium can be well described by a Coulomb potential with a linear potential as a perturbation, relevant for states close to the continuum. Their energy is approximately  given by $E_{n, \ell} = E_\text{Coulomb} + E_\text{Linear} \approx \alpha_{\rm eff}^2 M_\Q/4(-1/n^2 + 12 \epsilon n^2)$, with $\epsilon = \sigma/(\alpha_{\rm eff}^3 M_\Q^2) \approx \lambda_R \Lambda^2/(\alpha_{\rm eff}^3 M_\Q^2)$~\cite{Potential}. Here the potential part, dominated by the string tension $\sigma$, is approximated by $\lambda_R \Lambda^2$. The combination of the representations' quadratic Casimir operators $\lambda_R = 1/2 \left(2 C_{\rm adj. } - C_R\right)$ defines the strength of the channel. 

The above energy-conservation condition ($E_{n, \ell}  < \Lambda$) demands that the correction from the linear potential is not dominant and it thus limits the maximal partial wave $\ell$ by 
\begin{align}
& n_{\rm max} = \ell_{\rm max} + 1 = \left( \frac{\alpha_{\rm eff}^3 }{12\, \lambda_R}\right)^{1/4}\sqrt{M_\Q/ \Lambda} \nonumber \\
&= \frac{\alpha^{3/4} \lambda_R^{1/2}}{12^{1/4}}\sqrt{M_\Q/ \Lambda}\,.
\end{align}

After deriving the general bound $\ell_\text{max}$ from energy conservation, we come now to the momentum conservation condition. The maximal $\ell$ allowed by classical momentum conservation is
 \begin{align}
 \ell_{\rm class} =  p_Q \times R =  v_{\rm rel} M_\Q/\Lambda \approx  4 \sqrt{M_\Q T}/\Lambda\,, 
\end{align}
\noindent
where $1/\Lambda$ is the size of the $\Q g $ bound state. Since $\ell_{\rm class} $ scales with the temperature, there is a $T_c$ below which $\ell_{\rm class}< \ell_{\rm max}$, implying that all classically allowed partial waves contribute to the cross section, leading to a geometric interaction. Solving for this temperature, we get $T_c = \Lambda \, \lambda_R \, \alpha^{3/2} /8 \sqrt{3}$. This leads to a compact expression for the maximal partial wave allowed by energy conservation, $\ell_{\rm max}^2 =  4\, T_c  M_\Q/\Lambda^2$. 

The cross section for the reaction for the two quark state in representation R decomposed in partial waves reads 
\begin{align}
\sigma_{R}^{\rm ann} = p_R\, \,  4 \pi   \sum_{\ell=0}^{L_{\rm max}(T)}\frac{(2 \ell +1)\sin(\delta_\ell)^2 }{M_\Q^2 v_{\rm rel}^2}  \mathbb{P}_{\ell}^{\rm ann}\,,                                 
\end{align}
\noindent
where we define $L_{\rm max}(T) =\text{min}[\ell_{\rm max}, \ell_{\rm class}] $ which is set by either energy or momentum conservation respectively. $ \mathbb{P}_{\ell}^{\rm ann}$ is the probability that a bound state with given $\ell$ loses energy fast enough, falls to the ground state and self-annihilates. 

For this cross section, unitarity is saturated if all phases have $\sin(\delta_\ell) =1$. To derive the unitarity limit, it is sufficient to assume that all states that are allowed by energy and momentum conservation can self-annihilate fast enough and thus $\mathbb{P}_{\ell}^{\rm ann}=1$ for all of them. In fact, in Ref.~\cite{gluino}, it has been explicitly shown that this is close to the numerical result for the gluino. Furthermore, to obtain an upper bound on the cross section, the probability $p_R$ stemming from combinatorics can be set to unity as well. 

Summing the resulting cross sections up to the maximally allowed partial wave at a given temperature leads to the following maximal cross section compatible with unitarity for $T<\Lambda$:
\be\label{eq:suppressionfactor}
 (\sigma^{\rm ann}) \leq   \frac{4 \pi \,\ell_{\rm max}(T)^2 }{M_\Q^2 v_{\rm rel}^2} =  \sigma_{\rm geom.}  \left\{
\begin{array}{ll}
%0 & \mathrm{for}  \quad T>\Lambda\,, \\
 4 T_c/M_\Q v_{\rm rel}^2 & \mathrm{;} \, T > T_c,\\
1  & \mathrm{;} \, T<T_c\,,
\end{array}
\right.
 \ee
\noindent  with $\sigma_{\rm geom.} = 4 \pi/\Lambda^2$. This implies for the thermally averaged cross section,
\be\label{eq:suppressionfactor}
 \langle \sigma^{\rm ann} v_{\rm rel} \rangle  \leq 
  \sigma_{\rm geom.}  \, \sqrt{\frac{16 T}{\pi \, M_\Q}} \left\{
\begin{array}{ll}
%0 & \mathrm{for}  \quad T>\Lambda\,, \\
 T_c/T & \mathrm{for} \quad T > T_c,\\
1  & \mathrm{for} \quad  T<T_c\,.
\end{array}
\right.
\ee

It is obvious that the result of Ref.~\cite{GriestKamionkowski} is recovered at zero temperature, however, as in the perturbative calculation in the first part of this paper, we find that the effect of the thermal plasma reduces the cross section above the critical temperature $T_c$.  

The relic abundance is given by  
\begin{align}
& Y_{B} (\infty)
= \sqrt{\frac {45}{g_{\rm SM} \pi} }\frac 1 {M_\Q M_{\rm Pl}} \\ \nonumber
& \left( \frac{1}{Y_{B} (z_\Lambda)} + \int_{z=M_\Q/\Lambda}^{\infty} \frac{\langle \sigma^{\rm ann} v_{\rm rel} \rangle}{z^2} dz\right)^{-1}\,.
\end{align}
Integrating over both scaling regimes we get the asymptotic relic abundance
\begin{align}
&Y_{B} (\infty)= \frac{9 \sqrt{5\,M_\Q/g_{\rm SM}} }{8\, M_{\rm Pl}  \sigma_{\rm geom. }  \left( 3 \sqrt{\Lambda/T_c} -2\right) T_c^{3/2}  } \nonumber\\
& \approx \frac{3 \, \sqrt{5/g_{\rm SM}  }  }{8  M_{\rm Pl} \,\sigma_{\rm geom.  } T_c} \sqrt{\frac{M_\Q}{\Lambda}} \approx \frac{ 7.6\,  10^{-21}}{\rm GeV\, \lambda_R }\sqrt{\frac{ M_\Q \Lambda  }{\alpha^{3}}} \,. 
\end{align}

In fact, the relic density is dominantly determined at temperatures close to the phase transition, such that the geometrical cross section is less relevant for its asymptotic value than the cross section, which scales with temperature. This is a new result, obtained by taking into account the finite temperature effect and the basic principle of energy and momentum conservation.

\subsection{Implications of Geometrical Unitarity}
\begin{figure}[t]
%\begin{center}
\includegraphics[width=0.45\textwidth]{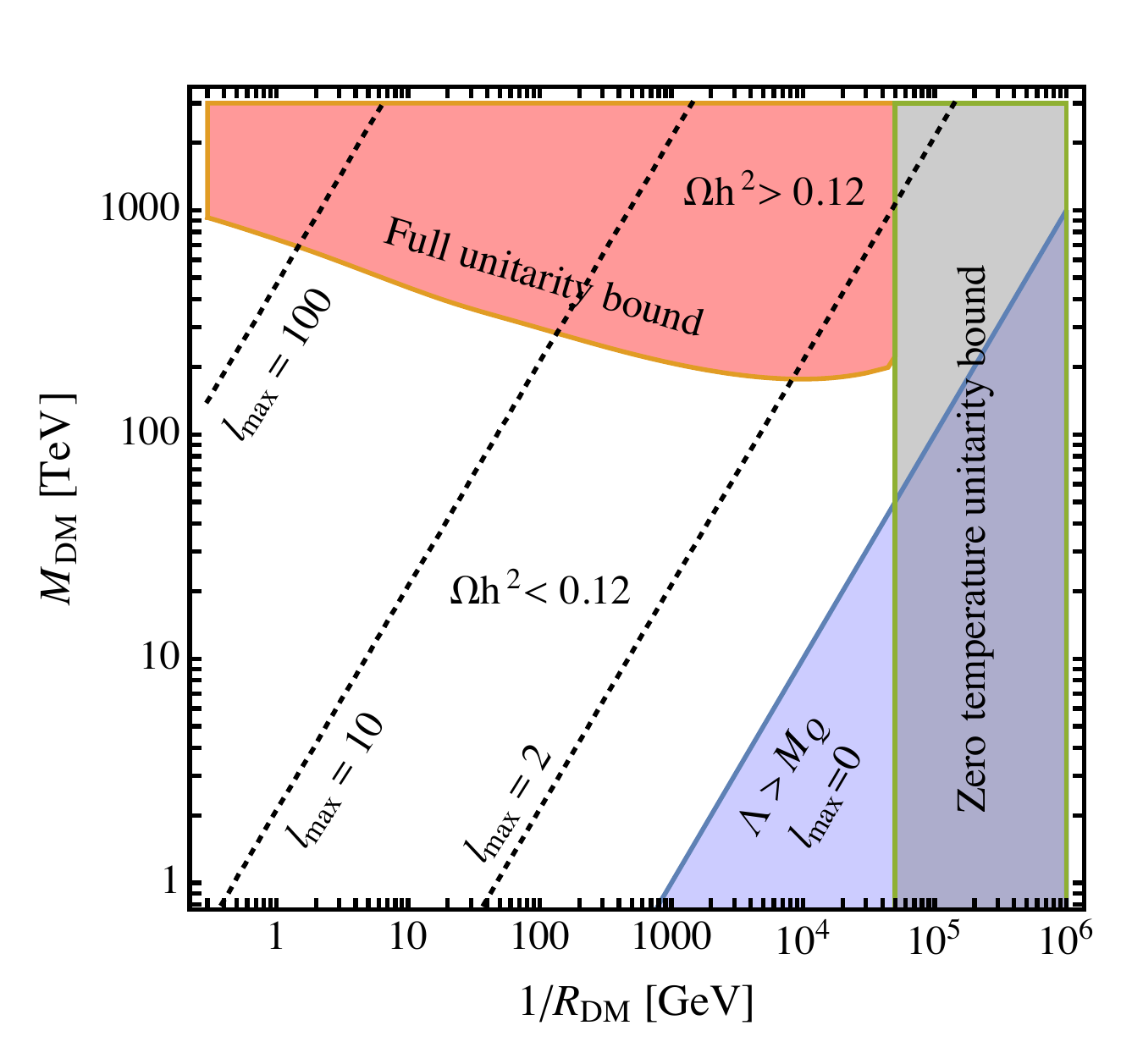}
\caption{\label{fig:RelicDGeo} The allowed parameter space for the simplest dark atoms (white area). The zero-temperature unitarity bound coincides with the full unitarity bound we derive only in the strong coupling regime. Geometrical annihilation (i.e., $\ell_{\rm max} \gg1$) only takes place in the regime where $M_\Q \gg \Lambda$. The values of $\ell_{\rm max} $ are indicated by dashed lines.}
%\end{center}
\end{figure}
\begin{figure}[t]
%\begin{center}
\includegraphics[width=0.45\textwidth]{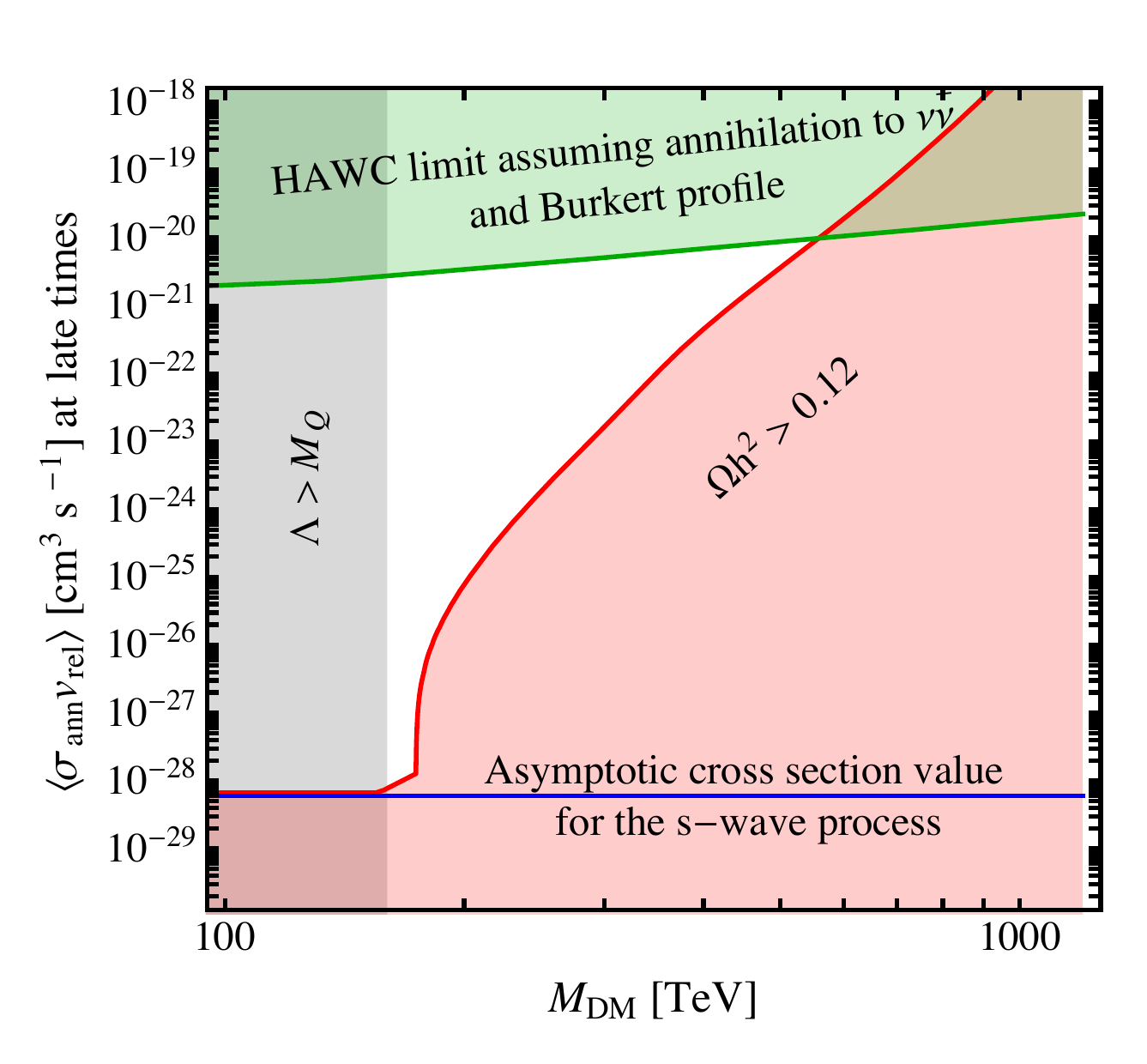}
\caption{\label{fig:LimitsGeo} The cross section for self-annihilation of dark atoms at late times. Superposed are limits of the HAWC experiment from Ref.~\cite{HAWC}, assuming complete annihilation to neutrino final states. At the large annihilation energies electro-weak corrections lead to gamma-ray emission, wich is constrained by the HAWC experiment.}
%\end{center}
\end{figure}
In Fig.~\ref{fig:RelicDGeo}, the results are summarized, with our bound coinciding  with the zero temperature bound of Ref.~\cite{GriestKamionkowski} only if $M_\Q \approx \Lambda$.  In this regime however, the dark atom mass is set by the scale $M_{\rm DM} \approx \Lambda$ and thus $R_{\rm DM } \approx 1/\Lambda \approx 1/M_{\rm DM}$, which implies that $\ell_{\rm max} = 0$. The reaction is therefore just an s-wave process and not geometrical because of the contributions of many partial waves.

On the contrary, if many partial waves contribute and $ \ell_{\rm max}  \gg 1$, we are in the regime with $M_\Q \gg \Lambda$ and $R_{\rm DM } \approx 1/\Lambda \gg 1/M_{\rm DM}$. We find that in this case the unitarity bound depends on the mass $M_\Q$ and results in a larger  DM radius corresponding to a monotonically decreasing $\Lambda$ with growing $M_\Q$. 

We demand that the scale $\Lambda > \Lambda_{\rm QCD}$, as otherwise big bang nucleosynthesis will be affected. The observation is that the existence of dark atoms would require a new confinement scale below approximately $60 \, \rm TeV$.  This contrasts with the finding of Ref.~\cite{GriestKamionkowski} that the limit on the DM size is independent of its mass. So, in the case of the lowest possible confinement scale $\Lambda \approx \Lambda_{\rm QCD}$, the DM mass can be as high as a PeV, but not much above that. Note that this upper bound on the mass also applies to a scenario where DM is co-annihilating with a partner particle, which has a non-abelian charge. For a detailed discussion, see Ref.~\cite{gluino}. 

In Fig.  \ref{fig:LimitsGeo}, the late time and low-temperature geometrical annihilation cross section of the dark atoms is shown. It exceeds by far the cross section expected from $s$-wave unitarity in the regime where $\Lambda \gg M_\Q$ and many partial waves contribute. This can be easily understood, as due to the effects of the plasma, the annihilation in the early universe is less efficient and the effective cross section setting the relic abundance is smaller given the same model parameters. 

When $\Lambda \approx M_\Q$ the cross section is $s$-wave dominated, but since the first step before the annihilation is a rearrangement process which is not strongly exothermic, we have $\sigma \approx \rm constant$ instead of $\sigma v \approx \rm constant$, and the annihilation is suppressed at late times. We note at this point that in this regime the quantum numbers of the DM particles with respect to the SM are important. When the geometrical annihilation cross section drops, they will become dominant and match the composite DM regime to the regime with elementary particles with dominant $s$-wave annihilation, as discussed in the first part of the paper. 

In Fig.  \ref{fig:LimitsGeo}, we compare the total annihilation cross section to the HAWC experiment bounds from Ref.~\cite{HAWC} assuming conservatively a dominant annihilation into neutrino final states, as suggested in Refs.~\cite{UpperBoundCross, NeutrinoConstraints, VHDM}, and a Burkert DM profile, which features a central core. The HAWC experiment can set limits on the annihilation cross section into neutrino final states, as at the large energies involved electro-weak corrections lead to unavoidable photon emission. Those corrections are taken into account by the PYTHIA software~\cite{1410.3012}, which was used in Ref.~\cite{HAWC}. Note that the predicted large annihilation cross sections at large DM masses are compatible with the excess of high energy neutrino events found in IceCube~\cite{IceCubeexcess}. We point out that composite dark matter, which saturates the unitarity can have intriguing direct detection signatures, as discussed in~\cite{1311.6386, 1505.02772}.

%%%%%%%%%%%%%%%%%%%%%%%%%%%%%%%%%%%%%%%%%%%%%%%%%%%%%%%%%%%%%%%%%%%%%
%                    Conclusions
%%%%%%%%%%%%%%%%%%%%%%%%%%%%%%%%%%%%%%%%%%%%%%%%%%%%%%%%%%%%%%%%%%%%%

\section{Conclusions} \label{sec:conclusions}

In searching to discover the particle nature of dark matter, the minimal thermal-relic WIMP model must be fully tested.  Because this model is defined by its annihilation cross section at freezeout in the early universe, the decisive test is to use searches for late-universe annihilation products to probe down to below the corresponding cross section scale.  Once this sensitivity is reached for a given mass, which requires allowing any combination of Standard-Model final states, then the minimal model is ruled out at that mass.  Importantly, the mass range is finite.  At the low-mass end, the thermal-WIMP hypothesis is being strongly tested (excluding neutrinos as final states, the current lower limit is about 20 GeV~\cite{LeaneSlatyerBeacom}), and that great progress is expected from new experiments.

But, to fully test thermal WIMPs, this hypothesis must also be strongly tested at high-mass end.  Here there are so far no experimental limits that approach the required sensitivity, so we must fall back on the unitarity bound, which intersects the thermal-relic cross section line in the range of hundreds of TeV.
 
Here we re-examine the physics of heavy WIMPs, focusing on a minimal model where the only new particle is the heavy WIMP itself so that its interactions must be mediated by Standard-Model particles.  Because those are relatively lighter than the heavy WIMP, they act as light mediators, which brings in Sommerfeld-enhancement effects, which are well studied, and bound-state effects, which are much less so.  We take advantage of the new formalism and results of Refs.~\cite{Petraki, Wise, PetrakiNew, CosmoBS, Tytgat}, which allow us to set up a formalism to study the freezeout of a particle in a generic $SU(2)_L$ representation, thus a true WIMP.  As the unitarity bound is approached, the growing interaction strength renders a tree-level approach inadequate.  Crucially, we use resummation techniques to take into account the non-perturbative effects of Sommerfeld enhancement and bound states.

For pointlike WIMPs, we find the following:
 
\begin{enumerate}

\item Taking into account bound-state formation effects in the early universe, the upper mass bound is reduced. In a UV-complete model of elementary dark-matter particles linked to the Standard Model by weak gauge bosons, the expected maximal dark-matter mass is lowered by about $30\%$. We find an upper bound of 144 TeV for self-conjugate dark matter and about 96 TeV for non-self-conjugate dark matter. Those mass scales, while high, are conceivably within reach of future collider experiments, more so then the oft-quoted value of $340$ [TeV]. 

\item The late-time annihilation of WIMPs is significantly enhanced due to Sommerfeld and bound-state formation effects. Furthermore, bound-state formation leads to new signals, such as capture photon emission and enhanced annihilation to monochromatic neutrinos.

We find that within the finite parameter space of the WIMP particle carrying $SU(2)_L$ charge, the direct detection signals are above the neutrino floor. Thus the entire parameter space will be tested by the next generation of underground direct-detection experiments. 

\end{enumerate}
 
For composite WIMPs, we find the following:
 
\begin{enumerate}

\item Taking into account the energy conservation of different partial waves contributing to geometrical annihilation, the resulting cross section cannot be constant at high temperatures. In particular, this finding leads to an upper bound on the dark matter radius, which depends on the dark matter mass, in sharp contrast to previous results~\cite{GriestKamionkowski}.  Further, demanding that the physics of the big bang nucleosynthesis remains unchanged, leads to an upper bound on the composite dark matter mass of about 1 PeV.

\item The newly found temperature dependence of the annihilation cross section is such that late-time annihilation is greatly enhanced. Therefore, despite the large mass, thermally produced extended dark matter particles appear to be testable by indirect detection experiments. Additionally, we find that in order to have a dark matter candidate near the heavy end of the allowed spectrum, the size of the composite dark matter grows, indicating a confining dynamics below the 100 TeV scale. 

\end{enumerate}
 
The phenomenological consequences of our framework are quite intriguing. We found that the bound-state formation processes at late times open up two new observational opportunities. On the one hand capture photons, which have the binding energy of the formed bound state and are nearly monochromatic are a new exciting signature for gamma-ray searches. On the other hand spin$-1$ bound states in our set-up have by $SU(2)_L$ symmetry significant annihilation rates to neutrinos and thus are good candidate processes to search for with experiments like IceCube, HAWC and Antares; see Refs.~\cite{IceCubeexcess, HAWC, Antares}.  In particular, the monochromatic nature of the neutrino signals allows a powerful background rejection of astrophysical sources and can become a unique tool for heavy dark matter searches. Concluding, we emphasise that the gamma-ray  (HAWC~\cite{1804.00628}, LHAASO~\cite{Ma:2018mzp}, CTA~\cite{1709.07997}), cosmic-ray (AMS~\cite{Zimmermann:2017abs}, CALET~\cite{Motz:2015qna}, DAMPE~\cite{TheDAMPE:2017dtc}), and high-energy neutrino experiments (IceCube~\cite{Flis:2017uay}, Antares~\cite{Antares}) provide a very promising route to test annihilations of TeV scale elementary and composite dark matter.

\vspace{0.4cm}

\section*{acknowledgments}

We thank Eric Braaten, Marc Kamionkowski, Mark Wise, Yue Zhang, and especially Iason Baldes, Simone Biondini, Marco Cirelli, Basudeb Dasgupta, Nicola Dondi, Ranjan Laha, Rebecca Leane, Kallia Petraki, Michele Redi, and  Bei Zhou for helpful comments on our manuscript.  

J.S. is thankful for the hospitality of CCAPP, where this project was initiated. The work of J.S. was supported by the INFN division in Florence and by the CP$^3$-Origins centre, which is partially funded by the Danish National Research Foundation, grant number DNRF90.  The work of J.F.B. was supported by NSF grant PHY-1714479.

%%%%%%%%%%%%%%%%%%%%%%%%%%%%%%%%%%%%%%%%%%%%%%%%%%%%%%%%
%%%%%%%%%%%%%%%%%%%%%%%%%%%%%%%%%%%%%%%%%%%%%%%%%%%%%%%%

\footnotesize
\bibliographystyle{abbrv}

\begin{thebibliography}{nnn}\bibitem{Ge:2019voa}
\heparticle[1903.05090]{S. Ge, K. Lawson, A. Zhitnitsky}{The Axion Quark Nugget DM Model: Size Distribution and Survival Pattern}


\bibitem{Berges:2019dgr}
\heparticle[1903.03116]{J. Berges, A. Chatrchyan, J. Jaeckel}{Foamy DM from Monodromies}


\bibitem{Blennow:2019fhy}
\heparticle[1903.00006]{M. Blennow, E. Fernandez-Martinez, A.O-D. Campo, S. Pascoli, S. Rosauro-Alcaraz, A.V. Titov}{Neutrino Portals to DM}


\bibitem{Arcadi:2019lka}
\heparticle[1903.03616]{G. Arcadi, A. Djouadi, M. Raidal}{DM through the Higgs portal}


\bibitem{Cirelli:2018iax}
\article[1811.03608]{M. Cirelli, Y. Gouttenoire, K. Petraki, F. Sala}{JCAP}{1902}{014}{2019}
{Homeopathic DM, or how diluted heavy substances produce high energy cosmic rays}


\bibitem{1705.05567}
\article[1705.05567]{B. Carr, M. Raidal, T. Tenkanen, V. Vaskonen, H. Veerm{\" a}e}{Phys. Rev.}{D96}{023514}{2017}
{Primordial black hole constraints for extended mass functions}


\bibitem{0404175}
\article[hep-ph/0404175]{G. Bertone, D. Hooper, J. Silk}{Phys. Rept.}{405}{279}{2004}
{Particle dark matter: Evidence, candidates and constraints}


\bibitem{Steigman:1984ac}
\article[Steigman:1984ac]{G. Steigman, M.S. Turner}{Nucl. Phys.}{B253}{375}{1985}
{Cosmological Constraints on the Properties of Weakly Interacting Massive Particles}


\bibitem{Arcadi:2017kky}
\article[1703.07364]{G. Arcadi, M. Dutra, P. Ghosh, M. Lindner, Y. Mambrini, M. Pierre, S. Profumo, F.S. Queiroz}{Eur. Phys. J.}{C78}{203}{2018}
{The waning of the WIMP? A review of models, searches, and constraints}


\bibitem{Roszkowski:2017nbc}
\article[1707.06277]{L. Roszkowski, E.M. Sessolo, S. Trojanowski}{Rept. Prog. Phys.}{81}{066201}{2018}
{WIMP DM candidates and searches—current status and future prospects}


\bibitem{SteigmanDasguptaBeacom}
\article[1204.3622]{G. Steigman, B. Dasgupta, J.F. Beacom}{Phys. Rev.}{D86}{023506}{2012}
{Precise Relic WIMP Abundance and its Impact on Searches for DM Annihilation}


\bibitem{Aghanim:2018eyx}
\heparticle[1807.06209]{{\sc Planck} Collaboration}{Planck 2018 results. VI. Cosmological parameters}


\bibitem{DM@LHC}
\article[1506.03116]{DM@LHC Working group}{Phys. Dark Univ.}{9-10}{8}{2016}
{Simplified Models for DM Searches at the LHC}


\bibitem{Baltz:2006fm}
\article[hep-ph/0602187]{E.A. Baltz, M. Battaglia, M.E. Peskin, T. Wizansky}{Phys. Rev.}{D74}{103521}{2006}
{Determination of dark matter properties at high-energy colliders}


\bibitem{Han:2018wus}
\article[1805.00015]{T. Han, S. Mukhopadhyay, X. Wang}{Phys. Rev.}{D98}{035026}{2018}
{Electroweak Dark Matter at Future Hadron Colliders}


\bibitem{Goodman:1984dc}
\article[Goodman:1984dc]{M.W. Goodman, E. Witten}{Phys. Rev.}{D31}{3059}{1984}
{Detectability of Certain Dark Matter Candidates}


\bibitem{Cushman:2013zza}
\heparticle[1310.8327]{Cushman et al.}{Working Group Report: WIMP Dark Matter Direct Detection}


\bibitem{Schumann:2019eaa}
\heparticle[1903.03026]{M. Schumann}{Direct Detection of WIMP Dark Matter: Concepts and Status}


\bibitem{UpperBoundCross}
\article[astro-ph/0608090]{J.F. Beacom, N.F. Bell, G.D. Mack}{Phys. Rev. Lett.}{99}{231301}{2006}
{General Upper Bound on the DM Total Annihilation Cross Section}


\bibitem{LeaneSlatyerBeacom}
\article[1805.10305]{R.K. Leane, T.R. Slatyer, J.F. Beacom, K.C.Y. Ng}{Phys. Rev.}{D98}{023016}{2018}
{GeV-scale thermal WIMPs: Not even slightly ruled out}


\bibitem{FermiDwarf}
\article[1601.06590]{{\sc MAGIC and Fermi-LAT} Collaborations}{JCAP}{1602}{039}{2016}
{Limits to DM Annihilation cross section from a Combined Analysis of MAGIC and Fermi-LAT Observations of Dwarf Satellite Galaxies}


\bibitem{NollettSteigman1}
\article[1312.5725]{K.M. Nollett, G. Steigman}{Phys. Rev.}{D89}{083508}{2014}
{BBN And The CMB Constrain Light, Electromagnetically Coupled WIMPs}


\bibitem{NollettSteigman2}
\article[1411.6005]{K.M. Nollett, G. Steigman}{Phys. Rev.}{D91}{083505}{2015}
{BBN And The CMB Constrain Neutrino Coupled Light WIMPs}


\bibitem{Galli:2013dna}
\article[1306.0563]{S. Galli, T.R. Slatyer, M. Valdes, F. Iocco}{Phys. Rev.}{D88}{063502}{2013}
{Systematic Uncertainties In Constraining Dark Matter Annihilation From The Cosmic Microwave Background}


\bibitem{Padmanabhan:2005es}
\article[astro-ph/0503486]{N. Padmanabhan, D.P. Finkbeiner}{Phys. Rev.}{D72}{023508}{2005}
{Detecting dark matter annihilation with CMB polarization: Signatures and experimental prospects}


\bibitem{1408.4131}
\article[1408.4131]{H. Silverwood, C. Weniger, P. Scott, G. Bertone}{JCAP}{1503}{055}{2015}
{A realistic assessment of the CTA sensitivity to dark matter annihilation}


\bibitem{GriestKamionkowski} 
\article[Griest:1989wd]{Kim Griest, Marc Kamionkowski}{\PRL}{64}{1990}{615}
{Unitarity Limits on the Mass and Radius of DM Particles}


\bibitem{Nima}
\article[0810.0713]{N. Arkani-Hamed, D.P. Finkbeiner, T.R. Slatyer, N. Weiner}{Phys. Rev.}{D79}{015014}{2008}
{A Theory of DM}


\bibitem{Cassel}
\article[0903.5307]{S. Cassel}{J. Phys.}{G37}{105009}{2009}
{Sommerfeld factor for arbitrary partial wave processes}


\bibitem{Petraki} 
\article[1407.7874]{B. von Harling, K. Petraki}{JCAP}{1412}{033}{2014}
{Bound-state formation for thermal relic DM and unitarity}


\bibitem{Wise} 
\article[1604.01776]{H. An, M.B. Wise, Y. Zhang}{Phys. Rev.}{D93}{115020}{2016}
{Effects of Bound States on DM Annihilation}


\bibitem{CosmoBS}
\article[1702.01141]{A. Mitridate, M. Redi, J. Smirnov, A. Strumia}{JCAP}{1705}{006}{2017}
{Cosmological Implications of DM Bound States}


\bibitem{Pavel}
\article[1409.8165]{M. Duerr, P. Fileviez Perez}{Phys. Rev.}{D91}{095001}{2015}
{Theory for Baryon Number and DM at the LHC}


\bibitem{DiracDM}
\article[1506.05107]{M. Duerr, P. Fileviez Perez, J. Smirnov}{Phys. Rev.}{D92}{083521}{2015}
{Simplified Dirac DM Models and Gamma-Ray Lines}


\bibitem{Hisano:2003ec}
\article[hep-ph/0307216]{J. Hisano, S. Matsumoto, M.M. Nojiri}{Phys. Rev. Lett.}{92}{031303}{2003}
{Explosive dark matter annihilation}.


\bibitem{Hisano:2004ds}
\article[hep-ph/0412403]{J. Hisano, S. Matsumoto, M.M. Nojiri, O. Saito}{Phys. Rev.}{D71}{063528}{2004}
{Non-perturbative effect on dark matter annihilation and gamma ray signature from galactic center}.


\bibitem{Hisano:2006nn}
\article[hep-ph/0610249]{J. Hisano, S. Matsumoto, M. Nagai, O. Saito, M. Senami}{Phys. Lett.}{B646}{34}{2006}
{Non-perturbative effect on thermal relic abundance of dark matter}.


\bibitem{0812.0559}
\article[0812.0559]{J.D. March-Russell, S.M. West}{Phys. Lett.}{B676}{133}{2008}
{WIMPonium and Boost Factors for Indirect DM Detection}


\bibitem{1609.00474}
\article[1609.00474]{S. Kim, M. Laine}{JCAP}{1701}{013}{2017}
{On thermal corrections to near-threshold annihilation}


\bibitem{1706.01894}
\article[1706.01894]{S. Biondini, M. Laine}{JHEP}{1708}{047}{2017}
{Re-derived overclosure bound for the inert doublet model}


\bibitem{PetrakiNew} 
\article[1612.07295]{M. Cirelli, P. Panci, K. Petraki, F. Sala, M. Taoso}{JCAP}{1705}{036}{2017}
{DM's secret liaisons: phenomenology of a dark U(1) sector with bound states}


\bibitem{Slatyer}
\article[1610.07617]{P. Asadi, M. Baumgart, P.J. Fitzpatrick, E. Krupczak, T.R. Slatyer}{JCAP}{1702}{005}{2017}
{Capture and Decay of Electroweak WIMPonium}


\bibitem{0901.2125}
\article[0901.2125]{W. Shepherd, T.M.P. Tait, G. Zaharijas}{Phys. Rev.}{D79}{055022}{2009}
{Bound states of weakly interacting DM}


\bibitem{Braaten1}
\article[1706.02253]{E. Braaten, E. Johnson, H. Zhang}{JHEP}{1711}{108}{2017}
{Zero-range effective field theory for resonant wino DM. Part I. Framework}


\bibitem{Braaten2}
\article[1708.07155]{E. Braaten, E. Johnson, H. Zhang}{JHEP}{1802}{150}{2018}
{Zero-range effective field theory for resonant wino DM. Part II. Coulomb resummation}


\bibitem{Braaten3}
\article[1712.07142]{E. Braaten, E. Johnson, H. Zhang}{JHEP}{1805}{062}{2018}
{Zero-range effective field theory for resonant wino DM. Part III. Annihilation effects}


\bibitem{PranNath}
\article[hep-ph/0702123]{D. Feldman, Z. Liu, P. Nath}{Phys. Rev.}{D75}{115001}{2007}
{The Stueckelberg Z-prime Extension with Kinetic Mixing and Milli-Charged DM From the Hidden Sector}


\bibitem{Ellis:2015vaa}
\article[1503.07142]{J. Ellis, F. Luo, K.A. Olive}{JHEP}{1509}{127}{2015}
{Gluino Coannihilation Revisited}


\bibitem{1703.00478}
\article[1703.00478]{I. Baldes, K. Petraki}{JCAP}{1709}{028}{2017}
{Asymmetric thermal-relic dark matter: Sommerfeld-enhanced freeze-out, annihilation signals and unitarity bounds}


\bibitem{HiggsMediatedBS}
\heparticle[1901.10030]{J. Harz, K. Petraki}{Higgs-mediated bound states in dark-matter models}


\bibitem{MinimalDM} 
\article[hep-ph/0512090]{M. Cirelli, N. Fornengo, A. Strumia}{Nucl. Phys.}{B753}{178}{2005}
{Minimal DM}


\bibitem{Feng:2010zp}
\article[1005.4678]{J.L. Feng, M. Kaplinghat, H-B. Yu}{Phys. Rev.}{D82}{083525}{2010}
{Sommerfeld Enhancements for Thermal Relic DM}


\bibitem{1611.04606}
\article[1611.04606]{A. Das, B. Dasgupta}{Phys. Rev. Lett.}{118}{251101}{2017}
{Selection Rule for Enhanced Dark Matter Annihilation}


\bibitem{1411.0752}
\article[1411.0752]{N. Nagata, S. Shirai}{Phys. Rev.}{D91}{055035}{2015}
{Electroweakly-Interacting Dirac Dark Matter}.


\bibitem{1805.01200}
\article[1805.01200]{J. Harz, K. Petraki}{JHEP}{1807}{096}{2018}
{Radiative bound-state formation in unbroken perturbative non-Abelian theories and implications for dark matter}


\bibitem{Kramer} 
\article{H. A. Kramers}{Philosophical Magazine,}{Series 6,}{1923}{Volume 46, Issue 275}{ On the theory of X-ray absorption and of the continuous X-ray spectrum}


\bibitem{Tytgat} 
\article[1711.08619]{L. Lopez Honorez, M.H.G. Tytgat, P. Tziveloglou, B. Zaldivar}{JHEP}{1804}{011}{2018}
{On Minimal DM coupled to the Higgs}


\bibitem{1611.01394}
\article[1611.01394]{K. Petraki, M. Postma, J. de Vries}{JHEP}{1704}{077}{2017}
{Radiative bound-state-formation cross sections for dark matter interacting via a Yukawa potential}


\bibitem{HiggsUnitarity}
\article[Lee:1977eg]{B.W. Lee, C. Quigg, H.B. Thacker}{Phys. Rev.}{D16}{1519}{1977}
{Weak Interactions at Very High-Energies: The Role of the Higgs Boson Mass}


\bibitem{FermiLAT}
\article[1601.06590]{{\sc Fermi-LAT } Collaboration}{JCAP}{1602}{039}{2016}
{Limits to DM Annihilation cross section from a Combined Analysis of MAGIC and Fermi-LAT Observations of Dwarf Satellite Galaxies}


\bibitem{1712.07489}
\article[1712.07489]{I. Baldes, M. Cirelli, P. Panci, K. Petraki, F. Sala, M. Taoso}{SciPost Phys.}{4}{041}{2018}
{Asymmetric dark matter: residual annihilations and self-interactions}


\bibitem{Bhanot:1979vb}
\article[Bhanot:1979vb]{G. Bhanot, M.E. Peskin}{Nucl. Phys.}{B156}{391}{1979}
{Short Distance Analysis for Heavy Quark Systems. 2. Applications}


\bibitem{1012.4515}
\article[1012.4515]{M. Cirelli, G. Corcella, A. Hektor, G. Hutsi, M. Kadastik, P. Panci, M. Raidal, F. Sala, A. Strumia}{JCAP}{1103}{051}{2010}
{PPPC 4 DM ID: A Poor Particle Physicist Cookbook for Dark Matter Indirect Detection}.


\bibitem{1712.07656}
\article[1712.07656]{M. Baumgart, T. Cohen, I. Moult, N.L. Rodd, T.R. Slatyer, M.P. Solon, I.W. Stewart, V. Vaidya}{JHEP}{1803}{117}{2018}
{Resummed Photon Spectra for WIMP Annihilation}


\bibitem{1808.08956}
\article[1808.08956]{M. Baumgart, T. Cohen, E. Moulin, I. Moult, L. Rinchiuso, N.L. Rodd, T.R. Slatyer, I.W. Stewart, V. Vaidya}{JHEP}{1901}{036}{2019}
{Precision Photon Spectra for Wino Annihilation}


\bibitem{0706.4071}
\article[0706.4071]{M. Cirelli, A. Strumia, M. Tamburini}{Nucl. Phys.}{B787}{152}{2007}
{Cosmology and Astrophysics of Minimal Dark Matter}.


\bibitem{1507.05519}
\article[1507.05519]{M. Cirelli, T. Hambye, P. Panci, F. Sala, M. Taoso}{JCAP}{1510}{026}{2015}
{Gamma ray tests of Minimal Dark Matter}.


\bibitem{1608.00786}
\article[1608.00786]{V. Lefranc, E. Moulin, P. Panci, F. Sala, J. Silk}{JCAP}{1609}{043}{2016}
{Dark Matter in $\gamma$ lines: Galactic Center vs dwarf galaxies}.


\bibitem{1810.00995}
\article[1810.00995]{{\sc H.E.S.S. } Collaboration}{JCAP}{1811}{037}{2018}
{Searches for gamma-ray lines and 'pure WIMP' spectra from Dark Matter annihilations in dwarf galaxies with H.E.S.S}.


\bibitem{LeptophillicDM}
\article[1401.6457]{J. Kopp, L. Michaels, J. Smirnov}{JCAP}{1404}{022}{2014}
{Loopy Constraints on Leptophilic DM and Internal Bremsstrahlung}


\bibitem{DirectVSIndirect} 
\article[1509.04282]{M. Duerr, P. Fileviez P{\' e}rez, J. Smirnov}{JHEP}{1606}{152}{2016}
{Scalar DM: Direct vs. Indirect Detection}


\bibitem{NuBoundsYuksel}
\article[0707.0196]{H. Yuksel, S. Horiuchi, J.F. Beacom, S'. Ando}{Phys. Rev.}{D76}{123506}{2007}
{Neutrino Constraints on the DM Total Annihilation Cross Section}


\bibitem{RunningJap} 
\article[1502.02244]{J. Hisano, R. Nagai, N. Nagata}{JHEP}{1505}{037}{2015}
{Effective Theories for DM Nucleon Scattering}


\bibitem{DARWIN}
\article[1606.07001]{{\sc DARWIN} Collaboration}{JCAP}{1611}{017}{2016}
{DARWIN: towards the ultimate dark matter detector}


\bibitem{XENON1T}
\article[1805.12562]{{\sc Xenon1T} Collaboration}{Phys. Rev. Lett.}{121}{111302}{2018}
{DM Search Results from a One Ton-Year Exposure of XENON1T}


\bibitem{BubbleSum}
\article[1708.00437]{G.M. Pelaggi, A.D. Plascencia, A. Salvio, F. Sannino, J. Smirnov, A. Strumia}{Phys. Rev.}{D97}{095013}{2018}
{Asymptotically Safe Standard Model Extensions?}


\bibitem{Holdom}
\article[1006.2119]{B. Holdom}{Phys. Lett.}{B694}{74}{2010}
{Large N flavor beta-functions: a recap}


\bibitem{1512.05353}
\article[1512.05353]{E. Del Nobile, M. Nardecchia, P. Panci}{JCAP}{1604}{048}{2016}
{Millicharge or Decay: A Critical Take on Minimal Dark Matter}.


\bibitem{Khlopov:2011tn}
\article[1111.2838]{M.Y. Khlopov}{Mod. Phys. Lett.}{A26}{2823}{2011}
{Physics of DM in the Light of Dark Atoms}


\bibitem{Khlopov:2014bia}
\article[1402.0181]{M.Y. Khlopov}{Int. J. Mod. Phys.}{A29}{1443002}{2014}
{Dark Atoms and Puzzles of DM Searches}


\bibitem{Tesi}
\heparticle[1812.08784]{M. Redi, A. Tesi}{Cosmological Production of Dark Nuclei}


\bibitem{gluino} 
\heparticle[1811.08418]{C. Gross, A. Mitridate, M. Redi, J. Smirnov, A. Strumia}{Cosmological Abundance of Colored Relics}


\bibitem{gluequark}
\heparticle[1811.06975]{R. Contino, A. Mitridate, A. Podo, M. Redi}{Gluequark DM}


\bibitem{Potential} 
\article[1801.01135]{V. De Luca, A. Mitridate, M. Redi, J. Smirnov, A. Strumia}{Phys. Rev.}{D97}{115024}{2018}
{Colored DM}


\bibitem{HAWC}
\article[1710.10288]{{\sc HAWC } Collaboration}{JCAP}{1802}{049}{2018}
{A Search for DM in the Galactic Halo with HAWC}


\bibitem{NeutrinoConstraints}
\article[0707.0196]{H. Yuksel, S. Horiuchi, J.F. Beacom, S'. Ando}{Phys. Rev.}{D76}{123506}{2007}
{Neutrino Constraints on the DM Total Annihilation Cross Section}


\bibitem{VHDM}
\article[1206.2595]{K. Murase, J.F. Beacom}{JCAP}{1210}{043}{2012}
{Constraining Very Heavy DM Using Diffuse Backgrounds of Neutrinos and Cascaded Gamma Rays}


\bibitem{1410.3012}
\article[1410.3012]{T. Sj{\" o}strand, S. Ask, J.R. Christiansen, R. Corke, N. Desai, P. Ilten, S. Mrenna, S. Prestel, C.O. Rasmussen, P.Z. Skands}{Comput. Phys. Commun.}{191}{159}{2015}
{An Introduction to PYTHIA 8.2}


\bibitem{IceCubeexcess} 
\article[1705.08103]{{\sc IceCube/DeepCore} Collaboration}{Eur. Phys. J.}{C77}{627}{2017}
{Search for Neutrinos from DM Self-Annihilations in the center of the Milky Way with 3 years of IceCube/DeepCore}


\bibitem{1311.6386}
\article[1311.6386]{R. Laha, E. Braaten}{Phys. Rev.}{D89}{103510}{2014}
{Direct detection of dark matter in universal bound states}


\bibitem{1505.02772}
\article[1505.02772]{R. Laha}{Phys. Rev.}{D92}{083509}{2015}
{Directional detection of dark matter in universal bound states}


\bibitem{Antares}
\article[1612.04595]{{\sc ANTARES } Collaboration}{Phys. Lett.}{B769}{249}{2017}
{Results from the search for DM in the Milky Way with 9 years of data of the ANTARES neutrino telescope}


\bibitem{1804.00628}
\article[1804.00628]{{\sc HAWC} Collaboration}{JCAP}{1806}{043}{2018}
{Search for Dark Matter Gamma-ray Emission from the Andromeda Galaxy with the High-Altitude Water Cherenkov Observatory}


\bibitem{Ma:2018mzp}
\article[Ma:2018mzp]{L. Ma}{PoS}{ICRC2017}{549}{2018}
{Expectation on Observation of Cosmic Rays Energy Spectrum from 10PeV to 100PeV with LHAASO Experiment}


\bibitem{1709.07997}
\heparticle[1709.07997]{{\sc CTA} Collaboration}{Science with the Cherenkov Telescope Array}


\bibitem{Zimmermann:2017abs}
\article[Zimmermann:2017abs]{N. Zimmermann}{PoS}{EPS-HEP2017}{090}{2017}
{Dark Matter signal from $\mathrm{e^{+}/e^{-}/\bar{p}}$ with the AMS-02 Detector on the International Space Station}


\bibitem{Motz:2015qna}
\article[Motz:2015qna]{H. Motz}{PoS}{ICRC2015}{1194}{2015}
{CALET's Sensitivity to Dark Matter and Astrophysical Sources}


\bibitem{TheDAMPE:2017dtc}
\article[1706.08453]{{\sc DAMPE} Collaboration}{Astropart. Phys.}{95}{6}{2017-10}
{The DArk Matter Particle Explorer mission}


\bibitem{Flis:2017uay}
\article[Flis:2017uay]{S. Flis, M. Medici}{PoS}{ICRC2017}{906}{2017}
{Searches for annihilating dark matter in the Milky Way halo with IceCube}


\end{thebibliography}

%=========================================================================

%%%%%%%%%%%%%%%%%%%%%%%%%%%%%%%%%%%%%%%%%%%%%%%%%%%%%%%%
%%%%%%%%%%%%%%%%%%%%%%%%%%%%%%%%%%%%%%%%%%%%%%%%%%%%%%%%

\end{document}